\documentclass{AIAA}

\usepackage[utf8]{inputenc}
\usepackage{latexsym}
\usepackage{amsmath}
\usepackage{amsfonts}

\usepackage{tikz}
\usetikzlibrary{patterns}
\usetikzlibrary{calc}
\usepackage{blindtext}
\usepackage{wrapfig}
\usepackage{graphicx}

\usepackage{tabularx}

\usepackage{multirow}

\newcommand{\twocell}[2][c]{
  \setlength{\extrarowheight}{-1em}
  \begin{tabular}[#1]{@{}c@{}}#2\end{tabular}}
 
\DeclareMathOperator*{\argmin}{arg\,min} 

\graphicspath{{./imgs/}{imgs/optimal_trajs/}{imgs/generalization/}{imgs/data_generation/}{imgs/networks/}{imgs/evaluation/}{imgs/rw/}{imgs/tv/}{imgs/simple/}}

\begin{document}

\title{Real-time optimal control via Deep Neural Networks: study on landing problems}

\author{Carlos S\'anchez-S\'anchez\footnote{Scientist, Advanced Concepts Team} and Dario Izzo\footnote{Scientific coordinator, Advanced Concepts Team.} }
\affiliation{European Space Agency, ESTEC, Noordwijk, The Netherlands.}

\begin{abstract}
Recent research on deep learning, a set of machine learning techniques able to learn deep architectures, has shown how robotic perception and action greatly benefits from these techniques. In terms of spacecraft navigation and control system, this suggests that deep architectures may be considered now to drive all or part of the on-board decision making system. In this paper this claim is investigated in more detail training deep artificial neural networks to represent the optimal control action during a pinpoint landing, assuming perfect state information. It is found to be possible to train deep networks for this purpose and that the resulting landings, driven by the trained networks, are close to simulated optimal ones. These results allow for the design of an on-board real time optimal control system able to cope with large sets of possible initial states while still producing an optimal response.

\end{abstract}

\maketitle

\section*{Nomenclature}
\noindent\begin{tabular}{@{}lcl@{}}
\textit{{$\mathbf x$}}          &=&     state \\
\textit{$\mathbf{r}$}          &=&     position vector \\
\textit{x}          &=&     horizontal position, m \\
\textit{y}          &=&     vertical position, m \\
\textit{$\mathbf{v}$}          &=&     velocity vector \\
\textit{$v_x$}          &=&     horizontal velocity, m/s \\
\textit{$v_y$}          &=&     vertical velocity, m/s \\
\textit{$\theta$}       &=&     pitch, rad \\
\textit{$m$}             &=&     mass, Kg \\
\textit{$\mathbf \lambda$}   &=&     costate vector \\
\textit{$\mathbf u$}       &=&     control variables\\
\textit{$\mathbf{g}$}          &=&     gravity force vector \\
\textit{g}          &=&     planetary gravity, m/s${}^2$ \\
\textit{$g_0$}          &=&     Earth's gravity, m/s${}^2$ \\
\textit{$I_{\text{sp}}$}       &=&     specific impulse, s\\
\textit{$\mathcal H$}   &=&     Hamiltonian \\
\textit{$J$}          &=&     cost function \\
\textit{$\gamma$}          &=&     weights of the cost function terms \\
\textit{$\mathcal A$}   &=&     initialization area for the spacecraft states\\
\textit{$\mathcal N$}   &=&     an artificial neural network \\
\textit{$\mathbf w, \mathbf b$}             &=& weights and biases of a neural network \\
\textit{$g$ }              &=&     activation function of a neural network
\end{tabular} 

\vskip 1cm

\textit{Subscripts}

\noindent\begin{tabular}{@{}lcl@{}}
\textit{$\text{QC}$}   &=&    quadratic control \\
\textit{$\text{MOC}$}   &=&     mass optimal control \\
\textit{$\text{T}$}   &=&     time optimal control \\
\textit{$0$}                &=&     initial\\
\textit{$t$}                &=&     target \\
\textit{$f$}                &=&     final \\
\end{tabular}

\section{Introduction}

Thanks to the decreasing cost of computational resources and to theoretical advances in research to train neural networks with many hidden~\cite{LeCun2015dlreview,schmidhuber2015dlreview} there is a renewed interest in artificial neural networks (ANN), and in particular in deep neural networks (DNN). DNNs are artificial neural networks with several hidden layers, that, layer after layer, form a hierarchical representation of an input-output map able, when correctly learned, to produce striking results. Examples of successful applications of DNNs include games AI \cite{silver2016go}, language processing \cite{collobert2011dnnnlp} and image understanding \cite{xu2015dnnimgunder} to just name some recent successes. 

While deep networks representation capabilities are particularly appropriate for perception related tasks such as image and speech recognition, it has been more recently pointed out how control problems may also benefit from these models \cite{ levine2013exploring, zhang15}. In these works, interesting results were obtained in cases where incomplete state information is available. Work in this direction has thus mostly taken the standpoint of dynamic programming, mapping the problem to that of learning some approximation of state-action pairs coming from a mixture of reinforcement learning techniques and dynamic programming algorithms. Instead, the use of deep artificial neural networks to approximate the state-action pairs computed from the solution to the optimal control problem of deterministic continuous non-linear systems, where the full state is directly observed, has been largely neglected. 

Some past attempts to explore possible uses of (shallow) ANNs in connection to the optimal control theory of deterministic continuous time non-linear systems are noteworthy, though limited to simple domains (e.g., linear systems often appearing in case studies) or to unbounded control \cite{effati2013optimal,xiong2014constrained,Medagam09}. Contributions to the solution of both the Hamilton-Jacobi-Belmann (HJB) equations and the two point boundary value problem resulting from Pontryagin's optimal control theory showed possible uses of ANNs in the domain of deterministic, continuous optimal control \cite{todorov2004optimality}. On the one hand, several methods were proposed for the approximation of the value function $v(t, \mathbf x)$ by means of ANNs architectures \cite{lewis2003hamilton,Tassa07,Medagam09}. On the other hand, ANNs have been proposed and studied to provide a trial solution to the states, to the co-states and to the controls so that their weights can be trained to make sure the assembled trial Hamiltonian respects Pontryagin's conditions \cite{effati2013optimal}. In this last case, the networks have to be retrained for each initial condition. Recursive networks have also been used to learn near-optimal controllers for motion tasks \cite{mordatch2015supervised} where the velocities (kinematics), and not the actual control, is to be predicted. A deep neural network trained with supervised signals, in the form of a stacked auto-encoder, was recently shown \cite{berniker2015deep} to be able to learn an accurate temporal profile of the optimal control and state in a point-to-point reach, non-linear limb model, but their architecture enforces the notable restriction of being based on a fixed-time problem. 

In this paper, DNNs are successfully trained to represent the solution to the Hamilton-Jacobi-Belmann policy equation in four different cases of pinpoint landing: a quadcopter model, a mass varying spacecraft with bounded thrust, a mass varying spacecraft equipped with a reaction wheel for attitude control and a mass varying rocket with thrust vector control. In all cases, the landing scenario is studied assuming perfect information on the spacecraft state. Approaches like guided policy search or dynamic programming hybrids \cite{levine2013exploring} are thus not necessary and a simpler training architecture can be assembled. Due to the assumptions considered in the models, feed-forward DNNs architectures can be trained directly in a supervised manner on the optimal state-action pairs obtained via an indirect method (based on single shooting). The trained networks are suitable for the on-board generation of descent guidance profiles as their computation requires a modest CPU effort. Training, on the other hand can be done offline and is thus not of concern to a real-time optimal control architecture.

The resulting DNNs thus enable real-time optimal control capabilities, without relying on optimal control methods (direct or indirect) on board, which could lead to an excessive use of the CPU and is undesirable due to numeric instabilities often connected to such solvers. In this sense, our work is related to previous attempts to obtain pinpoint landing guidance profiles computable on board \cite{acikmese2007convex}, and offers a novel, valid alternative. Remarkably, the learned policies have a validity which extends outside the area where training data is computed, contributing to their robustness and use possibilities. 

This paper builds on, and completes, previous work \cite{ICATT} where, notably, the state-action pairs was computed via direct methods and thus subject to chattering noise which prevented the study of more complex models such as pinpoint landing and thrust vectoring. The paper is structured as follows: in Section \ref{sec:optimalcontrol} we introduce the generic mathematical form of the optimal control problems (OCP) considered and we give the formal definition of the optimal control policy to be learned by the DNNs. In the following Section \ref{sec:optimalcontrolproblems}, four instances of OCPs are introduced, all related to pinpoint landing scenarios of relevance to aerospace systems, and, in each case, the two-point boundary value problem (TPBVP) is derived from the application of Pontryagin's maximum principle \cite{pontryagin1987mathematical}. In the following section \ref{sec:datageneration} it is described how the TPBVPs are solved by means of single shooting and continuation (homotopy) techniques as to generate training data (optimal state-action pairs) uniformly covering a large region of interest. In Section~\ref{sec:nnarchitecture} we describe the network architectures and training procedures used to approximate the optimal solutions. Section \ref{sec:evaluation} defines how the results of the DNNs are compared to the optimal trajectories and in Section \ref{sec:results} the performance of the networks is studied, including a comparison between different architectures and the study of the network behaviour for cases not considered in the training data.

\section{Optimal control}
\label{sec:optimalcontrol}
Lets consider deterministic systems defined by the time independent dynamics $\dot {\mathbf x}(t) = \mathbf f(\mathbf x(t), \mathbf u(t))$, where $\mathbf x(t): \mathbb R \rightarrow \mathbb R^{n_x}$ and $\mathbf u(t): \mathbb R \rightarrow \mathcal U \subset \mathbb R^{n_u}$. Consider the fundamental problem of finding an admissible control policy $\mathbf u(t)$ able to steer the system from any $\mathbf x_0 \in \mathbb R^{n_x}$ to some target $\mathcal S \subset \mathbb R^{n_x}$ in a (free) time $t_f$, while minimizing the cost function:
$$
J(\mathbf x(t), \mathbf u(t)) = \int_0^{t_f} \mathcal L(\mathbf x(t), \mathbf u(t)) dt + h(\mathbf x(t_f))
$$
The value function, defined as:
\begin{equation}
v(\mathbf x_0) = \min_{\mathbf u} J(\mathbf x(t), \mathbf u(t))
\label{eq:value}
\end{equation}
represents the minimal cost to reach the goal, starting from $\mathbf x_0$. Note how the value function is, in this case, not depending on time as $t_f$ is left free to be optimized. A finite horizon control problem is thus considered which has, mathematically, characteristics similar to an infinite horizon problem. Equivalently, the value function can be introduced as the solution to the partial differential equation \cite{todorov2004optimality}:


\begin{equation}
\min_{\mathbf u}\left\{ \mathcal L(\mathbf x, \mathbf u) + \mathbf f(\mathbf x, \mathbf u) \cdot \nabla_{\mathbf x} v(\mathbf x)\right\} = 0
\label{eq:hjb}
\end{equation}
subject to the boundary conditions $v(\mathbf x_t) = h(\mathbf x(t_f))$, $\forall \mathbf x_t \in \mathcal S$. The optimal control policy is then:
\begin{equation}
\mathbf u^*(\mathbf x) = \mbox{argmin}_{\mathbf u} \left\{ \mathcal L(\mathbf x, \mathbf u) + \mathbf f(\mathbf x, \mathbf u) \cdot \nabla_{\mathbf x} v(\mathbf x)\right\}
\label{eq:hjb2}
\end{equation}
Equations \ref{eq:hjb} and \ref{eq:hjb2} are the Hamilton-Jacobi-Bellman (HJB) equations for the free time, deterministic, optimal control problem here considered. They are a set of extremely challenging partial differential equations (PDEs) whose solution, pursued in the ``viscosity'' sense, is the solution to the original optimal control problem \cite{bardi2008optimal}. The HJB equations are important here as they imply the existence and uniqueness of an optimal state-feedback $\mathbf u^*(\mathbf x)$ which, in turn, allow to consider universal function approximators such as DNNs to represent it. Numerical approaches to solving HJB equations often rely on parametric approximations of the value function, e.g. using the Galerkin method \cite{beard1997galerkin}, and have thus also considered ANNs for the same purpose in the past \cite{Tassa07}. Here, deep neural networks (DNNs) are proposed to learn directly the optimal state-feedback $\mathbf u^*(\mathbf x)$ thus obtaining, indirectly, also a representation of the value function $v(\mathbf x) = J(\mathbf x^*, \mathbf u^*)$, while avoiding to make use of the network gradients when converting from value function to the optimal policy. Eventually, the trained DNN represents directly the optimal state-feedback and can be thus used, for example, in a non-linear model predictive control architecture \cite{izzo2013nonlinear} to achieve real-time optimal control capabilities.

\begin{table}
\caption{\label{table:control_problems} The four considered models at a glance.}
\begin{ruledtabular}
\begin{tabular}{lcccccc}
Model & $n_\mathbf{x}$ & $u_1$ & $u_2$ &  \twocell{Variable \\ mass}  & $\mathbf{g}$ &  \twocell{Optimization \\ problems} \\\hline
Quadcopter (QUAD) & 5 & N & rad/s & No & Earth & TOC, QC \\
Simple Sc. (SSC) & 4 & N & rad & Yes & Moon & MOC, QC \\
Reaction Wheel Sc. (RWSC) & 5 & N & rad/s & Yes & Moon & MOC, QC \\
Thrust Vectoring Rocket (TVR) & 6 & N & rad & Yes & Moon & MOC, QC \\
\end{tabular}
\end{ruledtabular}
\raggedright{\textit{QC: Quadratic control, TOC: Time-Optimal control, MOC: Mass-Optimal control. \\ $n_\mathbf{x}$: length of the state vector $\mathbf{x}$}}
\end{table}

\section{Optimal landing control problems}
\label{sec:optimalcontrolproblems}
\begin{figure}[t]
\centering
      \includegraphics[width=1\textwidth]{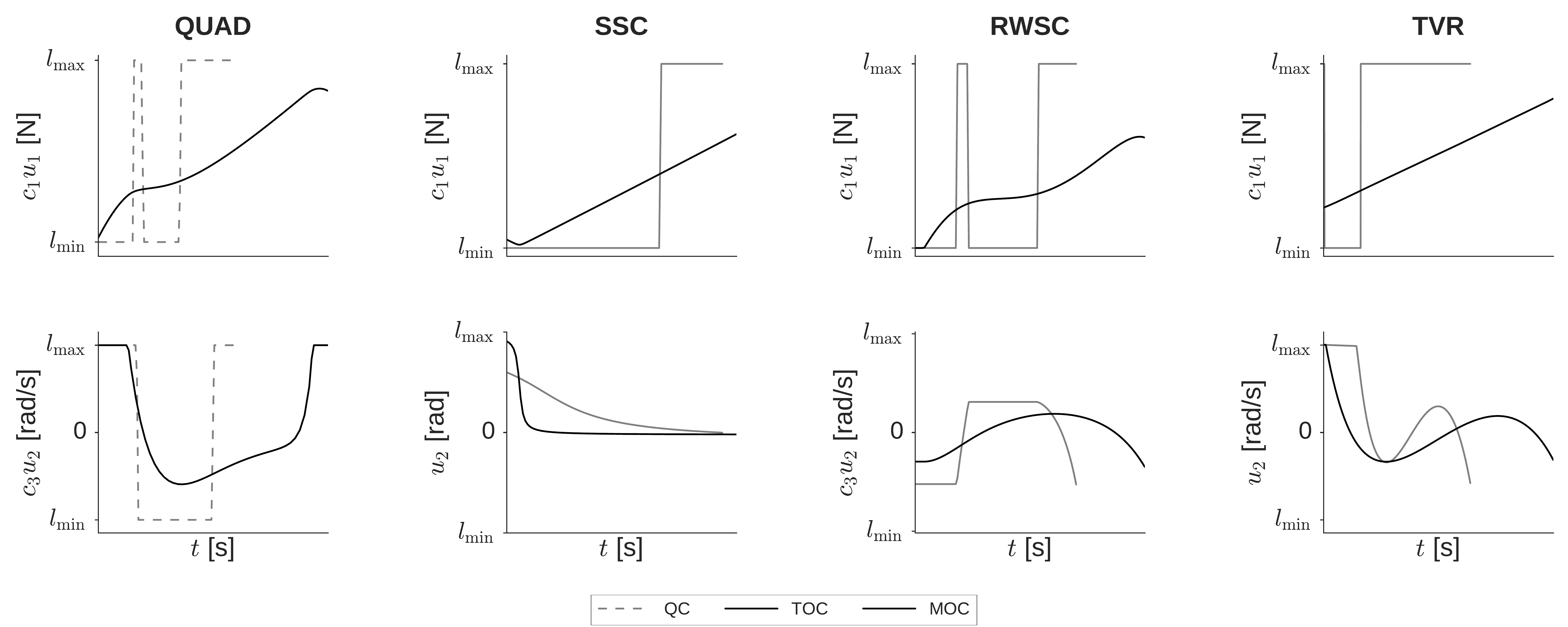} 
      \caption{Optimal control profiles of the models and objective functions here considered.}
      \label{fig:control_profiles} 
\end{figure}

The OCPs that are here considered correspond to different landing scenario all under a uniform gravity field, where the control is, in each case, defined by two variables $u_1, u_2$ ($n_u  =2$). Consider two different objectives: time optimal control (TOC) and quadratic control (QC) for the quadcopter model and mass optimal control (MOC) and quadratic control for the spacecraft models. The resulting set of test cases represent different classes of control profiles, as illustrated in Fig. \ref{fig:control_profiles}, including continuous control, discontinuous control, bang-off-bang control and saturated control. A summary of the models' characteristics is shown in Table \ref{table:control_problems}.

In the following subsections the details of each of the models considered are described and Pontryagin Maximum principle is used to derive the corresponding two point boundary value problem (TPBVP). If values for the initial values of the co-states and for the final time $t_f$ are found so that the dynamics and boundary conditions are satisfied as well as the additional condition $\mathcal H(t_f) = 0$ (a free time problem is considered), the corresponding control along the trajectory is assumed to be optimal and is used to create a number of optimal state-action pairs used for training the DNNs.

\subsection{Quadcopter}
Consider the following set of ordinary differential equations (ODE): 
\begin{equation}
\begin{array}{l}
    \dot{\mathbf r} = \mathbf v \\
    \dot{\mathbf v} = c_1 \frac {u_1}{m} \hat{\mathbf i}_\theta +  \mathbf g \\
    \dot{\theta} = c_2 u_2
\end{array}
\label{eq:Q_ODE}
\end{equation}
modelling the dynamics of a quadcopter moving in a two-dimensional space \cite{quadmodel}. The state is determined by the position $\mathbf r = (x,z)$, the velocity $\mathbf v = (v_x, v_z)$ and the orientation $\theta$ of the quadcopter. The mass of the quadcopter is $m = 1$ [Kg] and the acceleration due to the Earth's gravity is $\mathbf g = (0, -g)$ where $g=9.81$ [m/s$^2$]. The control $u_1 \in [0.05,1]$ models a thrust action applied along the direction ${\mathbf i}_\theta = [\sin \theta, \cos \theta]$ bounded by a maximum magnitude $c_1 = 20$ [N] and a minimum magnitude of $0.05c_1 = 1$ [N]. The control $u_2 \in [-1, 1]$ models the quadcopter pitch rate bounded by $c_2 = 2$ [rad/s]. Consider, as target state, $\mathbf r_t = (0,0)$, $\mathbf v_t = (0,0)$ and $\theta = 0$. Consider the minimization of the cost function: 
$$
J = (1 - \alpha) \int_0^{t_f} \left(\gamma_1 c_1^2u_1^2 + \gamma_2 c_2^2u_2^2\right) dt + \alpha \int_0^{t_f} dt
$$
where $\gamma_1 = 1$ [1/N$^2$] and $\gamma_2 = 1$ [s$^2$ / rad$^2$] are weights defining the balance between the cost of using $u_1$ or $u_2$ to control the quadcopter and $\alpha$ is a continuation parameter. The parameter $\alpha \in [0,1]$ defines a continuation between a quadratic optimal control problem (QC) $\alpha = 0$ and a time optimal control problem (TOC) $\alpha = 1$. Following Pontryagin \cite{pontryagin1987mathematical}, consider the following Hamiltonian:
$$
\mathcal H = \boldsymbol \lambda_{\mathbf r} \cdot \mathbf v + \boldsymbol \lambda_{\mathbf v} \cdot \left(c_1 u_1 \hat{\mathbf i}_\theta +  \mathbf g\right) + \lambda_\theta c_2u_2 + (1 - \alpha) (\gamma_1 c_1^2u_1^2 + \gamma_2 c_2^2u_2^2) + \alpha
$$
where the co-state functions $ \boldsymbol \lambda_{\mathbf r}(t), \boldsymbol \lambda_{\mathbf v}(t)$ and $\lambda_\theta(t)$ are introduced. Since $u_1\in[0.05, 1]$ and $u_2\in[-1, 1]$ both appear as a quadratic term, from the maximum principle it follows that their optimal values must be, if $\alpha\ne 1$:
\begin{equation}
\begin{array}{l}
u_1^* = \min\left(\max\left(-\frac{\boldsymbol \lambda_{\mathbf v} \cdot \hat{\mathbf i}_\theta }{2\gamma_1(1 - \alpha) c_1}, 0.05\right), 1\right)
\\
u_2^* = \min\left(\max\left(-\frac{\lambda_\theta }{2\gamma_2(1 - \alpha) c_2}, -1\right), 1\right)
\end{array}
\label{eq:quadp1}
\end{equation}
and, if $\alpha = 1$ (TOC case):
\begin{equation}
            u^*_1 = \left\{
            \begin{array}{ll}
                 1 & S_1<0  \\
                 0.05 & S_1>0
            \end{array}
            \right.\qquad
            u^*_2 = \left\{
            \begin{array}{ll}
                 1 & S_2<0  \\
                 -1 & S_2>0
            \end{array}
            \right.
\label{eq:quadp2}
\end{equation}
where $S_1 =  \boldsymbol \lambda_{\mathbf v} \cdot \hat{\mathbf i}_\theta$ and $S_2 = \lambda_\theta$ are called switching functions as they determine the control switch between extreme values of the domain where its defined. The differential equations defining the co-states ($\dot {\boldsymbol \lambda}_q = - \frac{\partial\mathcal H}{\partial q}$) are:
 \begin{equation}
\begin{array}{l}
    \dot{\boldsymbol \lambda}_{\mathbf r} = \mathbf 0 \\
    \dot{\boldsymbol \lambda}_{\mathbf v} = - {\boldsymbol \lambda}_{\mathbf r} \\
    \dot {\lambda}_\theta = - c_1 u_1 \boldsymbol \lambda_{\mathbf v} \cdot \hat{\mathbf i}_\tau
\end{array}
\label{eq:tv_costate}
\end{equation}
which can be simplified considerably as the first four differential equations are trivial. Overall the following holds:
\begin{equation}
\begin{array}{l}
    \lambda_x = \lambda_{x0} \\
    \lambda_z = \lambda_{z0} \\
    \lambda_{vx} = \lambda_{vx0} - \lambda_{x0} t \\
    \lambda_{vz} = \lambda_{vz0} - \lambda_{z0} t \\
    \dot {\lambda}_\theta = - c_1 u_1 [ (\lambda_{vx0} - \lambda_{x0} t) \cos\theta - (\lambda_{vz0} - \lambda_{z0} t) \sin\theta)]
\end{array}
\label{eq:tv_costate_simplified}
\end{equation}  
Eventually, the following two points boundary value problem (TPBVP) is obtained: 
\begin{equation}
\begin{array}{l}
    \dot{\mathbf r} = \mathbf v \\
    \dot{\mathbf v} = \frac{c_1 u_1^*} m \hat{\mathbf i}_\theta +  \mathbf g \\
    \dot \theta = c_2 u_2^* \\
    \dot {\lambda}_\theta = - c_1 u_1^* [ (\lambda_{vx0} - \lambda_{x0} t) \cos\theta - (\lambda_{vz0} - \lambda_{z0} t) \sin\theta)]
\end{array}
\label{eq:tv_tpbvp}
\end{equation}
with boundary conditions $\mathbf r_0, \mathbf v_0$, $\theta_0$ at $t=0$ and $\mathbf r_t = (0, 0), \mathbf v_t = (0, 0)$, $\theta_t = 0$ at $t=t_f$.

\label{sec:ssc}
\subsection{Simple spacecraft (SSC)}

Consider the following set of ODEs:

\begin{equation}
\begin{array}{l}
    \dot{\mathbf r} = \mathbf v \\
    \dot{\mathbf v} = c_1 \frac {u_1}m \hat{\mathbf i}_\theta +  \mathbf g \\
    \dot m = - \frac {c_1}{c_2} u_1
\end{array}
\label{eq:SSC_ODE}
\end{equation}
modelling the dynamics of a simple spacecraft in a two dimensional space (mass varying point mass). The state is determined by its position $\mathbf r = (x,z)$, its velocity $\mathbf v = (v_x, v_z)$ and its mass $m$. The acceleration $\mathbf g = (0, -g)$ considered is due to the Moon's gravity where $g = 1.6229$ [m/s$^2$]. The constant $c_2 = I_{sp} g_0$ represents the rocket engine efficiency in terms of its specific impulse $I_{sp} = 311$ [s] and $g_0 = 9.81$ [m/s$^2$]. The control $u_1 \in [0,1]$ models a thrust action applied along the direction ${\mathbf i}_\theta = [\sin \theta, \cos \theta]$ bounded by a maximum magnitude $c_1 = 44000$ [N]. Since the model does not include any rotational inertia, we assume to be able to freely steer the spacecraft pitch, so that $\theta$ is to be considered as a second control input $u_2$. Consider as a target the state $\mathbf r_t = (0,0)$, $\mathbf v_t = (0,0)$ and any $m$. Consider the minimization of the cost function 
$$
J = \frac{1}{c_2} \left[(1 - \alpha) \int_0^{t_f} \gamma_1 c_1^2 u_1^2 dt + \alpha \int_0^{t_f} c_1 u_1 dt\right]
$$
where $\gamma_1 = 1$ [1/N] is a weight defining a trade-off between the two contributions. The parameter $\alpha \in [0,1]$, defines a continuation between a quadratic optimal control problem (QC) $\alpha = 0$ and a mass optimal control problem (MOC) $\alpha = 1$. Following Pontryagin \cite{pontryagin1987mathematical}, consider the following Hamiltonian:
$$
\mathcal H = \boldsymbol \lambda_{\mathbf r} \cdot \mathbf v + \boldsymbol \lambda_{\mathbf v} \cdot \left(c_1 \frac {u_1}m \hat{\mathbf i}_\theta +  \mathbf g\right) - \lambda_m \frac {c_1}{c_2} u_1 +  \frac{1}{c_2} \left[(1 - \alpha) \gamma_1 c_1^2 u_1^2 + \alpha c_1 u_1 \right]
$$
where the co-state functions $ \boldsymbol \lambda_{\mathbf r}(t), \boldsymbol \lambda_{\mathbf v}(t)$ and $\lambda_m(t)$ are introduced. From the maximum principle it immediately follows that, necessarily, the optimal value for $u_2$ (and hence $\theta$), must be:
$$
\hat{\mathbf i}_\theta^* = -\frac{\boldsymbol \lambda_{\mathbf v}}{\lambda_v}
$$
while for $u_1$, since it appears quadratically we may conclude, if $\alpha\ne 1$:
$$
u_1^* = \min\left(\max\left(\frac{\frac{\lambda_v c_2}{m}+\lambda_m-\alpha}{2\gamma_1c_1(1-\alpha)}, 0\right), 1\right)
$$
and, if $\alpha = 1$ (MOC case):
\begin{equation*}
u^*_1 = \left\{
    \begin{array}{ll}
        1 & S_1<0  \\
        0 & S_1>0
    \end{array}
    \right.
\label{eq:ssc2}
\end{equation*}
where $S_1 = \alpha - \frac{\lambda_v c_2}{m} - \lambda_m$ is the switching function for this problem. The differential equations defining the co-states ($\dot {\boldsymbol \lambda}_q = - \frac{\partial\mathcal H}{\partial q}$) are:
\begin{equation}
\begin{array}{l}
    \dot{\boldsymbol \lambda}_{\mathbf r} = \mathbf 0 \\
    \dot{\boldsymbol \lambda}_{\mathbf v} = - {\boldsymbol \lambda}_{\mathbf r} \\
    \dot \lambda_m = \frac{c_1}{m^2} {\boldsymbol \lambda}_{\mathbf v} \cdot \hat{\mathbf i}_\theta u
\end{array}
\label{eq:SSC_costate}
\end{equation}
\noindent
Eventually, the following two points boundary value problem (TPBVP) is obtained: 
\begin{equation}
\begin{array}{l}
    \dot{\mathbf r} = \mathbf v \\
    \dot{\mathbf v} = c_1 \frac {u_1^*}m \hat{\mathbf i}_\theta^* +  \mathbf g \\
    \dot m = - \frac {c_1}{c_2} u_1^* \\
    \dot {\lambda}_m = \frac {c_1}{m^2} u_1^* [ (\lambda_{vx0} - \lambda_{x0} t) \sin\theta + (\lambda_{vz0} - \lambda_{z0} t) \cos\theta)]
\end{array}
\label{eq:SSC_tpbvp}
\end{equation}
with boundary conditions $\mathbf r_0, \mathbf v_0$, $m_0$ at $t=0$ and $\mathbf r_t = (0, 0), \mathbf v_t = (0, 0)$ and $\lambda_{mt} = 0$ at $t=t_f$.

\begin{figure}[t]
\centering
      \includegraphics[width=0.48\textwidth]{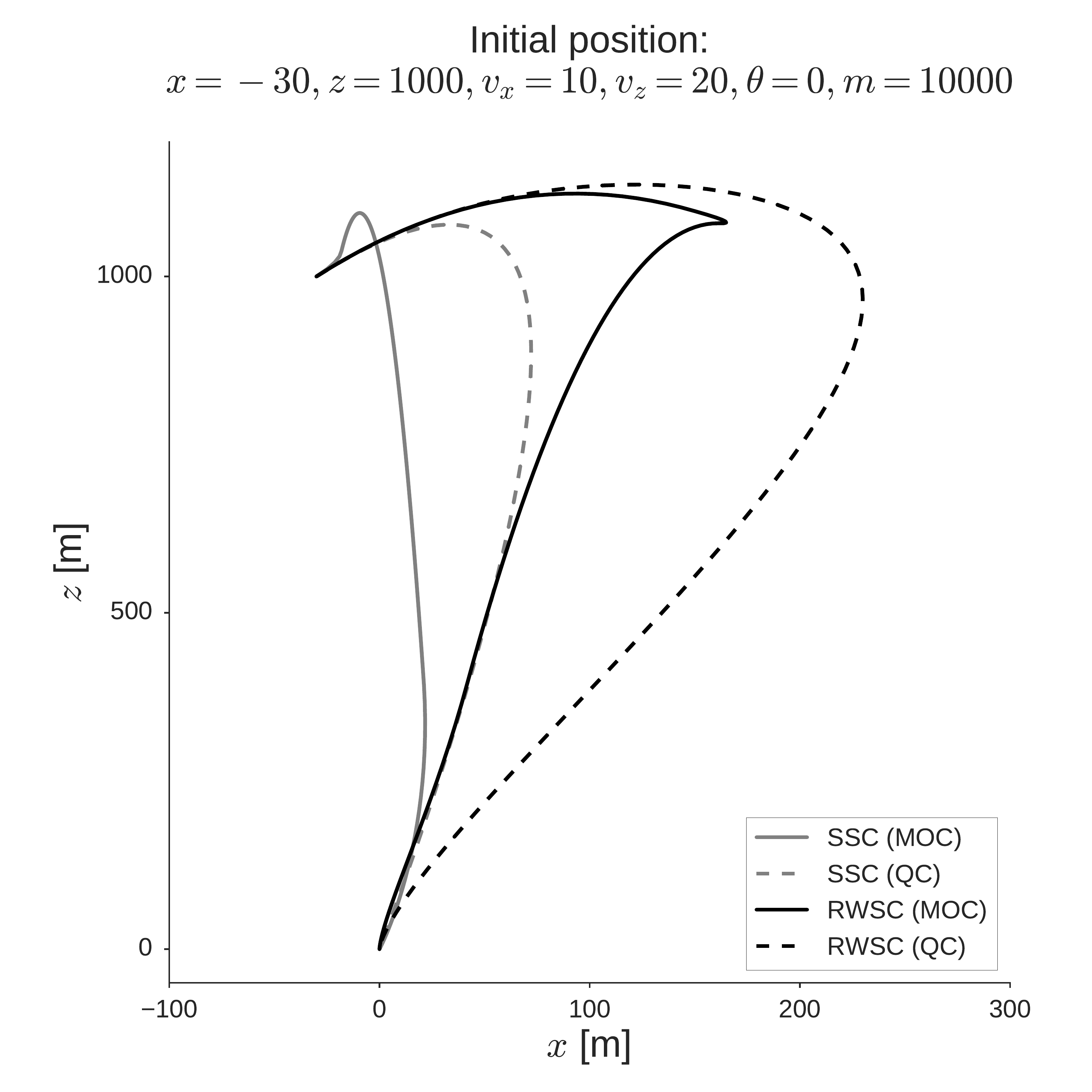} 
      \caption{Trajectories from the same initial state for the two spacecraft models and the two objective functions.}
      \label{fig:spacecraft_trajs} 
\end{figure}

\subsection{Reaction wheel spacecraft (RWSC)}

In this model, the spacecraft attitude is controlled by a reaction wheel able to induce a bounded angular velocity on the spacecraft body (we neglect the fact that, as the spacecraft becomes lighter, the maximum angular velocity also increases, as well as the fact that the wheel may get saturated). The state is, thus, the position $\mathbf r = (x,z)$, the velocity $\mathbf v = (v_x, v_z)$ and the mass $m$. The system dynamics is described by the following set of ODEs:

\begin{equation}
\begin{array}{l}
    \dot{\mathbf r} = \mathbf v \\
    \dot{\mathbf v} = c_1 \frac {u_1} {m} \hat{\mathbf i}_\theta +  \mathbf g \\
    \dot\theta = c_3 u_2 \\
    \dot m = - \frac {c_1}{c_2} u_1
\end{array}
\label{eq:rw_state}
\end{equation}
Similarly to what was considered for the SSC model, the acceleration $\mathbf g = (0, -g)$ is due to the Moon's gravity where $g=1.6229$ [m/s$^2$], while the constant $c_2 = I_{sp} g_0$ represents the rocket engine efficiency in terms of its specific impulse $I_{sp} = 311$ [s] and $g_0 = 9.81$ [m/s$^2$]. While the control $u_1$ has the same meaning as in the previous SSC model, $u_2 \in [-1,1]$ now corresponds to the pitch rate control actuated by a reaction wheel and is bounded by a maximum magnitude of $c_3=0.0698$ [rad/s]. This difference also results in rather different optimal landing trajectories as illustrated in the example in Figure \ref{fig:spacecraft_trajs}. Consider as a target state for this system $\mathbf r_t = (0,0)$, $\mathbf v_t = (0,0)$, $\theta_t = 0$ and $m = \mbox{any}$ and note that, unlike in the SSC case, now a terminal vertical descent is forced thanks to the final condition on the attitude $\theta_t = 0$. Consider the minimization of the cost function:
$$
J = (1-\alpha) \int_0^{t_f}\left[\frac{\gamma_1 c_1^2}{c_2} u_1^2 + c_3^2 u_2^2 \right]dt + \alpha \int_0^{t_f}\left[\frac{\gamma_2 c_1}{c_2} u_1 + c_3^2 u_2^2 \right] dt
$$ 
where $\gamma_1 = 1.5E^{-6}$ [(rad$^2$ s) / (kg$^2$ m)] and $\gamma_2 = 1.5E^{-2}$ [(rad$^2$) / (s kg)] are defining the cost trade-off between the use of $u_1$ and $u_2$ to control the spacecraft trajectory and are chosen as to give priority to optimize the use of $u_1$ (i.e. the thruster). The parameter $\alpha \in [0,1]$ defines a continuation between a quadratic optimal control problem (QC, $\alpha = 0$) and a mass optimal control problem (MOC, $\alpha = 1$) for $u_1$, while for $u_2$ the power spent by the reaction wheel is always considered. Following Pontryagin \cite{pontryagin1987mathematical}, consider the following Hamiltonian:
$$
\mathcal H = \boldsymbol \lambda_{\mathbf r} \cdot \mathbf v + \boldsymbol \lambda_{\mathbf v} \cdot \left(c_1 \frac {u_1}{m} \hat{\mathbf i}_\theta +  \mathbf g\right) - \lambda_m \frac {c_1}{c_2} u_1 + \lambda_\theta c_3u_2 + (1-\alpha)\frac{\gamma_1 c_1^2}{c_2} u_1^2  + \alpha\frac{\gamma_2 c_1}{c_2} u_1 + c_3^2 u_2^2
$$
where the co-state functions $ \boldsymbol \lambda_{\mathbf r}(t), \boldsymbol \lambda_{\mathbf v}(t)$, $\lambda_m(t)$ and $\lambda_\theta$ are introduced. From the maximum principle it immediately follows that, necessarily, the optimal value for the controls, must be, if $\alpha\ne 1$:
\begin{equation}
\begin{array}{l}
u_1^* = \min\left(\max\left(-\frac{\boldsymbol \lambda_{\mathbf v} \cdot \hat{\mathbf i}_\theta \frac{c_2}{m} - \lambda_m + \alpha \gamma_2 }{2(1-\alpha)\gamma_1 c_1}, 0\right), 1\right)
\\
u_2^* = \min\left(\max\left(-\frac{\lambda_\theta }{2c_3}, -1\right), 1\right)
\end{array}
\label{eq:rwsc1}
\end{equation}
and, if $\alpha = 1$ (MOC case):
\begin{equation*}
u^*_1 = \left\{
    \begin{array}{ll}
        1 & S_1<0  \\
        0 & S_1>0
    \end{array}
    \right.
\label{eq:rwsc2}
\end{equation*}
where $S_1 = \boldsymbol \lambda_{\mathbf v} \cdot \hat{\mathbf i}_\theta \frac{c_2}{m} - \lambda_m + \alpha \gamma_2$ is the switching function for this problem. The differential equations defining the co-states ($\dot {\boldsymbol \lambda}_q = - \frac{\partial\mathcal H}{\partial q}$) are:
\begin{equation}
\begin{array}{l}
    \dot{\boldsymbol \lambda}_{\mathbf r} = \mathbf 0 \\
    \dot{\boldsymbol \lambda}_{\mathbf v} = - {\boldsymbol \lambda}_{\mathbf r} \\
    \dot {\lambda_\theta} = - \frac{c_1}{m} {\boldsymbol \lambda}_{\mathbf v} \cdot \hat{\mathbf i}_\tau u_1 \\
    \dot \lambda_m =  \frac{c_1}{m^2} {\boldsymbol \lambda}_{\mathbf v} \cdot \hat{\mathbf i}_\theta u_1
\end{array}
\label{eq:rw_costate}
\end{equation}
where we have introduced the unit vector $\hat{\mathbf i}_\tau = [\cos\theta, -\sin\theta]$. Eventually, the following two points boundary value problem (TPBVP) is obtained: 
\begin{equation}
\begin{array}{l}
    \dot{\mathbf r} = \mathbf v \\
    \dot{\mathbf v} = c_1 \frac {u_1} {m} \hat{\mathbf i}_\theta +  \mathbf g \\
    \dot\theta = c_3 u_2 \\
    \dot m = - \frac {c_1}{c_2} u_1 \\
    \dot {\lambda}_\theta = - \frac {c_1}{m} u_1^* [ (\lambda_{vx0} - \lambda_{x0} t) \cos\theta - (\lambda_{vz0} - \lambda_{z0} t) \sin\theta)]\\
    \dot {\lambda}_m = \frac {c_1}{m^2} u_1^* [ (\lambda_{vx0} - \lambda_{x0} t) \sin\theta + (\lambda_{vz0} - \lambda_{z0} t) \cos\theta)]
\end{array}
\label{eq:rwsc_tpbvp}
\end{equation}
with boundary conditions $\mathbf r_0, \mathbf v_0$, $m_0$, $\theta_0$ at $t=0$ and $\mathbf r_t = (0, 0), \mathbf v_t = (0, 0)$ and $\lambda_{mt} = 0$ at $t=t_f$.

\subsection{Thrust vectoring rocket (TVR)}

\begin{figure}[t!]
\centering
\begin{tikzpicture}
    \draw[thin,->] (0,0) -- (3,0) node[anchor=north west] {$\hat{\mathbf i}_x$};
    \draw[thin,->] (0,0) -- (0,3) node[anchor=south east] {$\hat{\mathbf i}_y$};
    
    \def\rocketrot{30}
    \begin{scope}[rotate=-\rocketrot]
        \def\x{0}
        \def\y{0}
        \def\w{0.2}
        \def\h{3}

        \coordinate (LL) at (\x - \w / 2, \y - \h / 2);
        \coordinate (LR) at (\x + \w / 2, \y - \h / 2);
        \coordinate (UL) at (\x - \w / 2, \y + \h / 2);
        \coordinate (UR) at (\x + \w / 2, \y + \h / 2);
        \coordinate (NOSE) at (\x, \y + \h / 2 + \h / 6);
        \coordinate (Torigin) at (\x,\y - \h / 2 - \h / 12);
        \coordinate (Tvect) at (1.5,1.5);
        \coordinate (Norigin) at (\x, \y - \h / 2);
        \coordinate (NL) at (\x - \w / 2, \y - \h / 2 - \h / 12);
        \coordinate (NR) at (\x + \w / 2, \y - \h / 2 - \h / 12);
        
        \coordinate (O) at (\x, \y + \h / 2 + \h / 3);
        
        \draw[thin,->] (O) -- ($(O) + (1,0)$) node[anchor=north west] {$\hat{\mathbf i}_\tau$};
        \draw[thin,->] (O) -- ($(O) + (0,1)$) node[anchor=south west] {$\hat{\mathbf i}_\theta$};
        
        \fill[fill=black!70!white](0,0) -- (\w/2, 0) arc (0:90:\w/2);
        \fill[fill=black!70!white](0,0) -- (-\w/2, 0) arc (180:270:\w/2);
        \draw (0,0) circle (\w/2);
        
        \draw[] (LL) -- (UL) -- (NOSE) -- (UR) -- (LR) -- (LL);
        
        \draw[] (Norigin) -- (NL) -- (NR) -- (Norigin);
        
        \draw[thin, dashed] (0,-2) -- (0,3);
        
        \draw [thick, ->] (Torigin) -- ($(Torigin) + (Tvect)$) node[midway,below] {$c_1 u_1 \hat{\mathbf t}$};
    \end{scope}
    
    \draw [thick, ->] (0,0) -- (0,-1) node[midway,right] {$\mathbf g$};
    
    \draw (0,0) ++(90:.8) arc (90:90-\rocketrot:.8) node[midway,above] {$\theta$};
\end{tikzpicture}
\caption{The thrust vectoring model.\label{fig:tv}}
\end{figure}
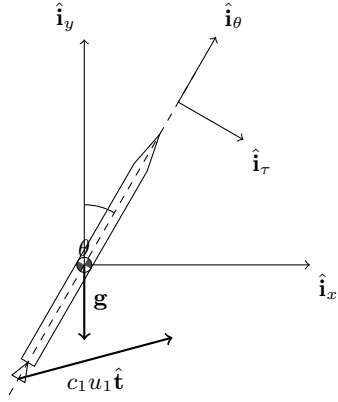

Consider the following set of ODE:
\begin{equation}
\begin{array}{l}
    \dot{\mathbf r} = \mathbf v \\
    \dot{\mathbf v} = c_1 \frac {u_1} {m} \hat{\mathbf t} +  \mathbf g \\
    \dot \theta = \omega \\
    \dot \omega = - c_1 \frac {u_1} {Rm} \hat{\mathbf t} \cdot \hat{\mathbf i}_\tau \\
    \dot m = - \frac {c_1}{c_2} u_1
\end{array}
\label{eq:tv_odes}
\end{equation}
modelling the dynamics of a rocket moving in a two-dimensional space and controlled by thrust vectoring as illustrated in Fig.\ref{fig:tv}.  The state is determined by position $\mathbf r = (x,z)$, velocity $\mathbf v = (v_x, v_z)$, orientation $\theta$, angular velocity $\omega$ and mass $m$ of the rocket. The acceleration due to Moon's gravity is $\mathbf g = (0, -g)$ where $g = 1.6229$ [m/s$^2$]. The constant $c_2 = I_{sp} g_0$ represents the rocket engine efficiency in terms of its specific impulse $I_{sp} = 311$ [s] and $g_0 = 9.81$ [m/s$^2$]. The control $u_1 \in [0,1]$ [N] models a thrust action applied along the direction $\hat{\mathbf t}$ and bounded by a maximum magnitude $c_1 = 20$ [N]. The control $u_2 \in [-\phi, \phi]$ models the thrust vector tilt with respect to the symmetry axis, so that $\hat{\mathbf t} = \cos(\theta+u_2) \hat{\mathbf i}_x+ \sin(\theta+u_2)\hat{\mathbf i}_y$. We will consider $\phi = 10 [deg]$ here. Consider, as target state, $\mathbf r_t = (0,0)$, $\mathbf v_t = (0,0)$, $\theta_t = 0$ and $\omega_t = 0$. Consider the minimization of the cost function (note that the cost is the same as that defined for the SSC case in Section \ref{sec:ssc}): 
$$
J = \frac{1}{c_2} \left[(1 - \alpha) \int_0^{t_f} \gamma_1 c_1^2 u_1^2 dt + \alpha \int_0^{t_f} c_1 u_1 dt\right]
$$
where $\gamma_1 = 1$ [1/N] is a weight defining a trade-off between the two contributions. The parameter $\alpha \in [0,1]$, defines a continuation between a quadratic optimal control problem (QC) $\alpha = 0$ and a mass optimal control problem (MOC) $\alpha = 1$. Following Pontryagin \cite{pontryagin1987mathematical}, consider the following Hamiltonian:

\begin{equation}
    \mathcal H = \boldsymbol \lambda_{\mathbf r} \cdot \mathbf v + \boldsymbol \lambda_{\mathbf v} \cdot \left(c_1 \frac {u_1} {m} \hat{\mathbf t} +  \mathbf g\right) + \lambda_\theta \omega - \lambda_\omega c_1\frac {u_1} {Rm} \hat{\mathbf t} \cdot \hat{\mathbf i}_\tau - \lambda_m \frac {c_1}{c_2} u_1 + \frac{1 - \alpha}{c_2}\gamma_1 c_1^2 u_1^2 + \frac{c_1}{c_2} \alpha u_1
\end{equation}
From the maximum principle it immediately follows that, necessarily, the optimal value for $\hat{\mathbf t}$  (and hence $u_2$), must be:
$$
\hat{\mathbf t}^* = - \frac{\boldsymbol \lambda_{\mathbf v} -  \frac{\lambda_\omega}R \hat{\mathbf i}_\tau}{|\boldsymbol \lambda_{\mathbf v} -  \frac{\lambda_\omega}R \hat{\mathbf i}_\tau|} = - \frac{\boldsymbol \lambda_{aux}}{\lambda_{aux}}
$$
where we have introduced the auxiliary co-state $\boldsymbol\lambda_{aux} = \boldsymbol \lambda_{\mathbf v} -  \frac{\lambda_\omega} R \hat{\mathbf i}_\tau $. The Hamiltonian along an optimal trajectory may be then rewritten as:
\begin{equation}
    \mathcal H = \boldsymbol \lambda_{\mathbf r} \cdot \mathbf v + \boldsymbol \lambda_{\mathbf v} \cdot \mathbf g + \lambda_\theta \omega - \lambda_m \frac {c_1}{c_2} u_1 - \lambda_{aux} \frac{c_1}{m} u_1 
    + \frac{1 - \alpha}{c_2}\gamma_1 c_1^2 u_1^2 + \frac{c_1}{c_2} \alpha u_1
\end{equation}
and, since $u_1$ appears as a quadratic term, its optimal value must be, if $\alpha\ne 1$:
$$
u_1^* = \min\left(\max\left(\frac{\lambda_m + \frac{c_2}{m} \lambda_{aux}
    - \alpha}{2\gamma_1 (1-\alpha)c_1}, 0\right), 1\right)
$$
and, if $\alpha = 1$ (MOC case):
\begin{equation*}
u^*_1 = \left\{
    \begin{array}{ll}
        1 & S_1<0  \\
        0 & S_1>0
    \end{array}
    \right.
\label{eq:TVR2}
\end{equation*}
where $S_1 =  \alpha - \lambda_m - \frac{c_2}{m} \lambda_{aux}$ is the switching function for this problem. The differential equations for the costates ($\dot \lambda_q = - \frac{\partial\mathcal H}{\partial q}$) are:
 \begin{equation}
\begin{array}{l}
    \dot{\boldsymbol \lambda}_{\mathbf r} = \mathbf 0 \\
    \dot{\boldsymbol \lambda}_{\mathbf v} = - {\boldsymbol \lambda}_{\mathbf r} \\
    \dot {\lambda}_\theta = - \frac{\lambda_\omega}R c_1 \frac {u_1}m \hat{\mathbf t} \cdot \hat{\mathbf i}_\theta \\
    \dot{\lambda}_\omega = - \lambda_\theta \\ 
    \dot{\lambda}_m = \frac{c_1 u_1}{m^2} (\boldsymbol \lambda_{\mathbf v} -  \frac{\lambda_\omega} R \hat{\mathbf i}_\tau) \cdot \hat{\mathbf t} 
\end{array}
\label{fig:tv_costate}
\end{equation}
Eventually, the following two points boundary value problem (TPBVP) is obtained: 
\begin{equation}
\begin{array}{l}
    \dot{\mathbf r} = \mathbf v \\
    \dot{\mathbf v} = c_1 \frac {u_1^*} {m} \hat{\mathbf t}^* +  \mathbf g \\
    \dot \theta = \omega \\
    \dot \omega = - c_1 \frac {u_1^*} {Rm} \hat{\mathbf t}^* \cdot \hat{\mathbf i}_\tau \\
    \dot m = - \frac {c_1}{c_2} u_1^*\\
    \dot {\lambda}_\theta = - \frac{\lambda_\omega}R c_1 \frac {u_1^*}m \hat{\mathbf t^*} \cdot \hat{\mathbf i}_\theta \\
    \dot{\lambda}_\omega = - \lambda_\theta \\ 
    \dot{\lambda}_m = \frac{c_1 u_1^*}{m^2} (\boldsymbol \lambda_{\mathbf v} -  \frac{\lambda_\omega} R \hat{\mathbf i}_\tau) \cdot \hat{\mathbf t}^*
\end{array}
\label{eq:tv_state}
\end{equation}
where $\boldsymbol \lambda_{\mathbf v} = [\lambda_{vx0} + \lambda_{x0}t, \lambda_{vz0} + \lambda_{z0}t]$ and with boundary conditions $\mathbf r_0, \mathbf v_0$, $\theta_0$, $\omega_0$ at $t=0$ and $\mathbf r_t = (0, 0), \mathbf v_t = (0, 0)$, $\theta_t = 0$, $\omega_t = 0$ and $\lambda_{mt} = 0$ at $t=t_f$.

\section{Generating the training and validation data}
\label{sec:datageneration}

A dataset containing optimal trajectories is generated for each of the problems described in the previous section. Each optimal trajectory consists of a list of pairs $(\mathbf x^*, \mathbf u^*)$ where $\mathbf x^*$ is the state and $\mathbf u^*$ is the corresponding optimal action. For each one of the problems, an initialization area $\mathcal A$ is defined to draw the initial conditions from, so that, formally, $x_0 \in \mathcal A$. The definition of $\mathcal A$ for each model can be found in table \ref{table:bounds}. $135,000$ optimal trajectories are generated for each problem, from each optimal trajectory $100$ state-control pairs are uniformly selected along the trajectory and inserted in the training data (thus containing 13,500,000 optimal state action pairs). 

The direct method used in previous work to compute the optimal trajectories \cite{ICATT} resulted in chattering problems due to numeric instabilities, while the profiles obtained via the indirect methods used in this paper are clean and accurately represent the optimal control without the need for extra regularization, as illustrated by Figure \ref{fig:direct_indirect}. The direct method, in this case, produces a control with some chattering having a minor overall effect on the predicted optimal trajectory, but posing a major problem if the state action pairs have to be used in a training set.

\begin{figure}[t]
\centering
      \includegraphics[width=0.8\textwidth]{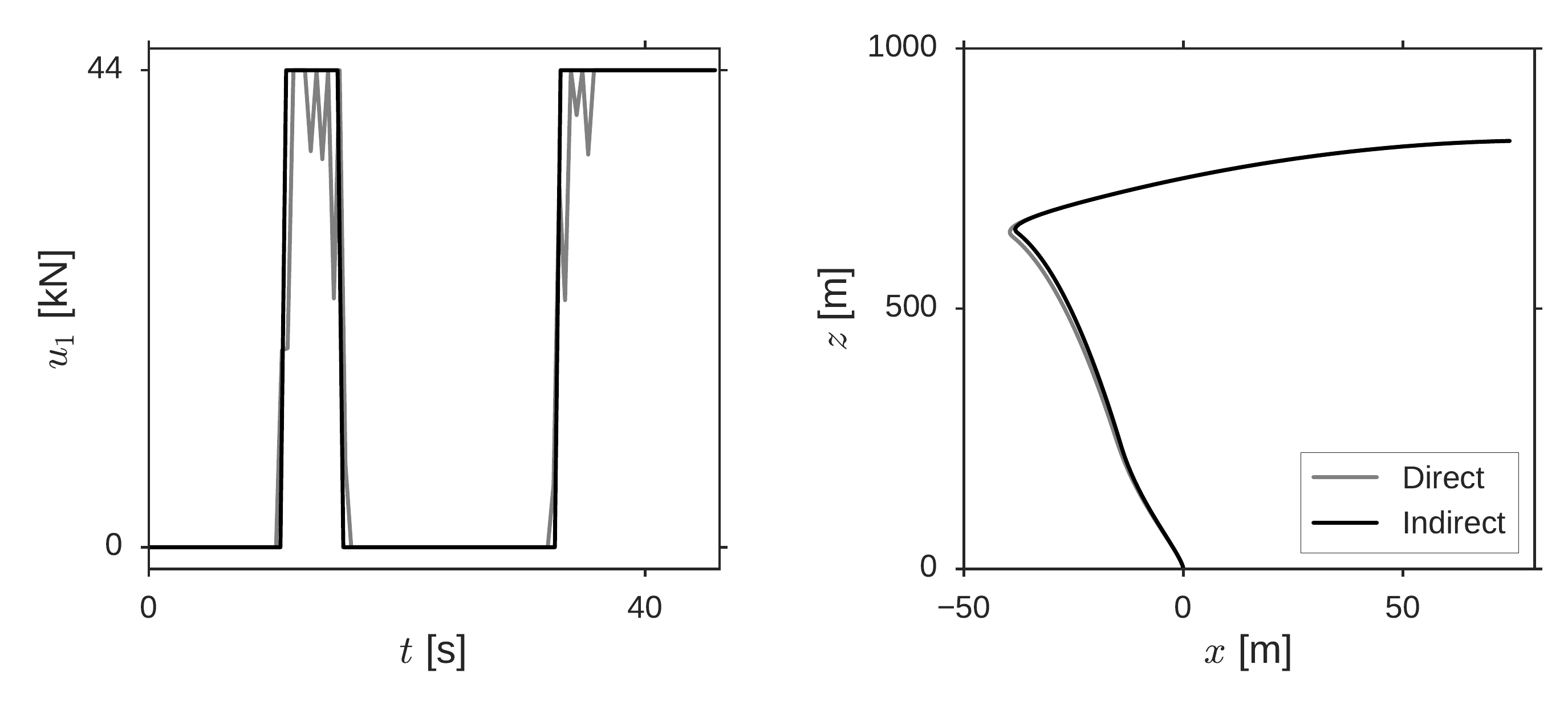} 
      \caption{RWSC-MOC problem solved using direct and indirect methods. Thrust control (left) and trajectory (right)}
      \label{fig:direct_indirect} 
\end{figure}

For the single shooting method to converge, the initial guess for the costates $\boldsymbol \lambda_0$ is required to be close to the optimal solution. The quadratic control problems, given the smoothness of their solutions, can be solved from a random initial guess inside some relatively broad bounds. However, for the time and mass optimal control cases (TOC and MOC, both corresponding to $\alpha = 1$), a more precise guess is needed. Continuation methods are thus here used to reduce the required computation time in the former and to be able to find the solutions in the latter, thus using as the initial guess for $\boldsymbol \lambda_0$ the solution to an optimal problem that is expected to be similar.

\begin{table}[]
\centering
\caption{Initialization areas $\mathcal A$ for the four landing problems.  \label{table:bounds}}

\begin{ruledtabular}
\begin{tabular}{lcccccccc}

& $x$ [m] & $z$ [m] &  $v_x$ m/s & $v_z$ [m/s] & $\theta$ [rad] & $\omega$ [rad/s] & $m$ [kg] \\ \hline

\textbf{QUAD} &  [-5, 5] & [2.5,20] & [-1, 1] & [-1,1] & $\left [- \frac{\pi}{10}, \frac{\pi}{10} \right ]$ & - & -\\

\textbf{SSC} & [-200, 200] & [500,2000] &  [-10, 10] & [-30,10] & - & - & [8000, 12000]\\

\textbf{RWSC} & [-200, 200] & [500,2000] & [-10, 10] & [-30,10] &  $\left [- \frac{\pi}{20}, \frac{\pi}{20} \right]$  & - & [8000, 12000]\\

\textbf{TVR}  & [-10, 10] & [500,2000] & [-0.5, 0.5] & [-40,0] &  $\left[-10^{-3}, 10^{-3}\right]$ & $\left[-10^{-4}, 10^{-4}\right]$ & [8000, 12000]\\
\\

\end{tabular}

\end{ruledtabular}

\end{table}

\begin{figure}[t]
\centering
      \includegraphics[width=1\textwidth]{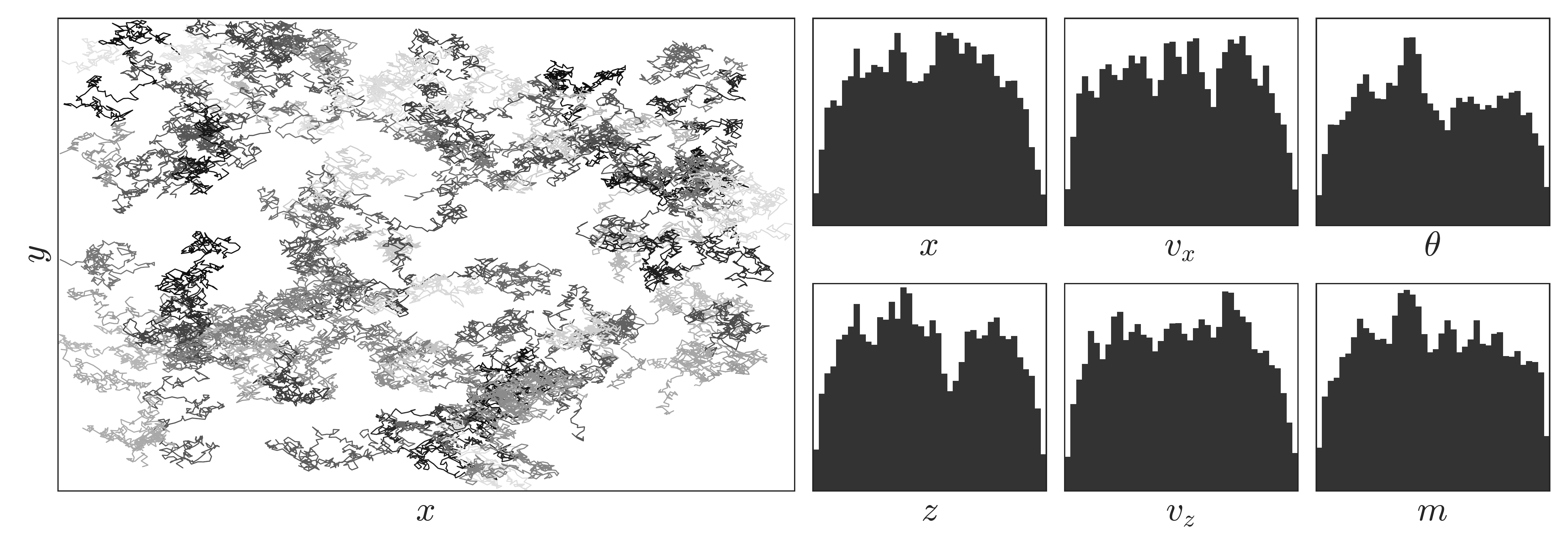} 
      \caption{Location of initial states generated by 100 random walks and distribution of initial states (RWSC).}
      \label{fig:random_walks} 
\end{figure}

To find the TOC or MOC solution, the corresponding quadratic control problem is solved first and then an homotopy path is followed by continuously increasing $\alpha$ from $0$ to $1$, resulting in smooth changes as shown by Figure \ref{fig:homotopy}. Once a solution (QC, TOC or MOC) with an associated costate vector $\boldsymbol \lambda_0$ is found for an initial  state $\boldsymbol x_0$, a random walk in the $n_x$ dimensional state space is initiated. The costate vector  $\boldsymbol \lambda_0$ will be used as the initial guess to find the optimal trajectory starting from a point $\boldsymbol x_0'$ generated by perturbing each variable $x_i \in \boldsymbol x_0 $ so that $x_i' =  x_i + \delta$ with $\delta$ being a step size drawn from the uniform distribution $\delta \sim U(0, \eta r_i$), where  $r_i$ corresponds to the range of the initialization area for the variable $x_i$ and $\eta = 0.02$ determines the maximum step size. The random walk continues until reaching the bounds of $\mathcal A$ or $300$ trajectories have been generated to then start a new random walk. Eventually, enough optimal trajectories are found to form the evaluation set. Separate and independent random walks are then started to create the validation set, up to when 15,000 optimal trajectories and 1,500,000 optimal state action pairs are found.
\begin{figure}[t]
\centering
      \includegraphics[width=0.5\textwidth]{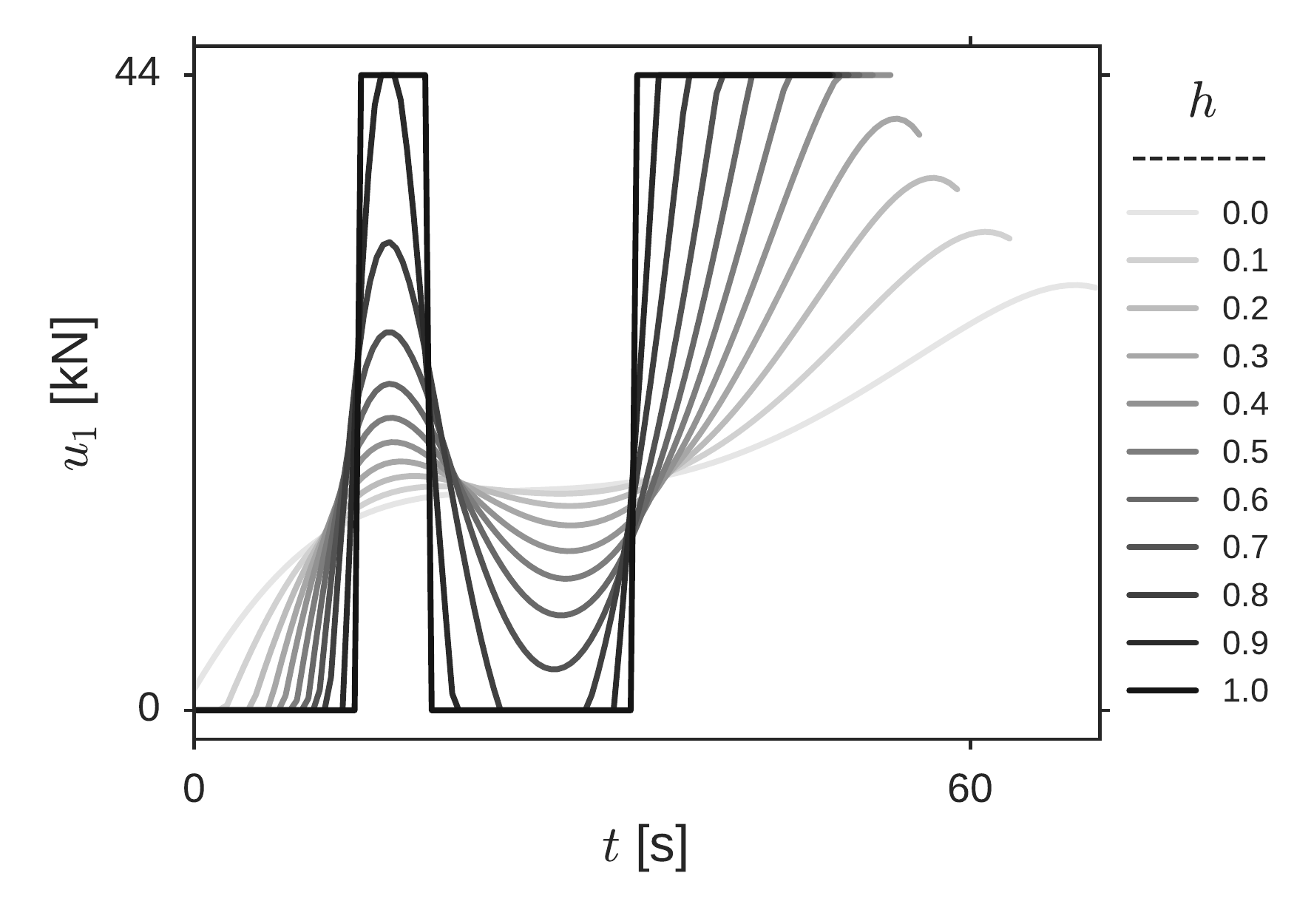} 
      \caption{Continuation from quadratic control ($h=0$) to mass optimal control ($h=1$) for the RWSC model.}
      \label{fig:homotopy} 
\end{figure}
For the quadrotor, the simple and the reaction wheel spacecrafts models, the initial state initializing each random walk is uniformly drawn at random in $\mathcal A$ resulting in a quite uniform coverage of the space $\mathcal A$. Figure \ref{fig:random_walks} shows some of the random walks as well as the distribution of the initial states for the RWSC-MOC problem, showing that this method achieves an uniform coverage of $\mathcal{A}$

\begin{figure}[t]
\centering
      \includegraphics[width=1\textwidth]{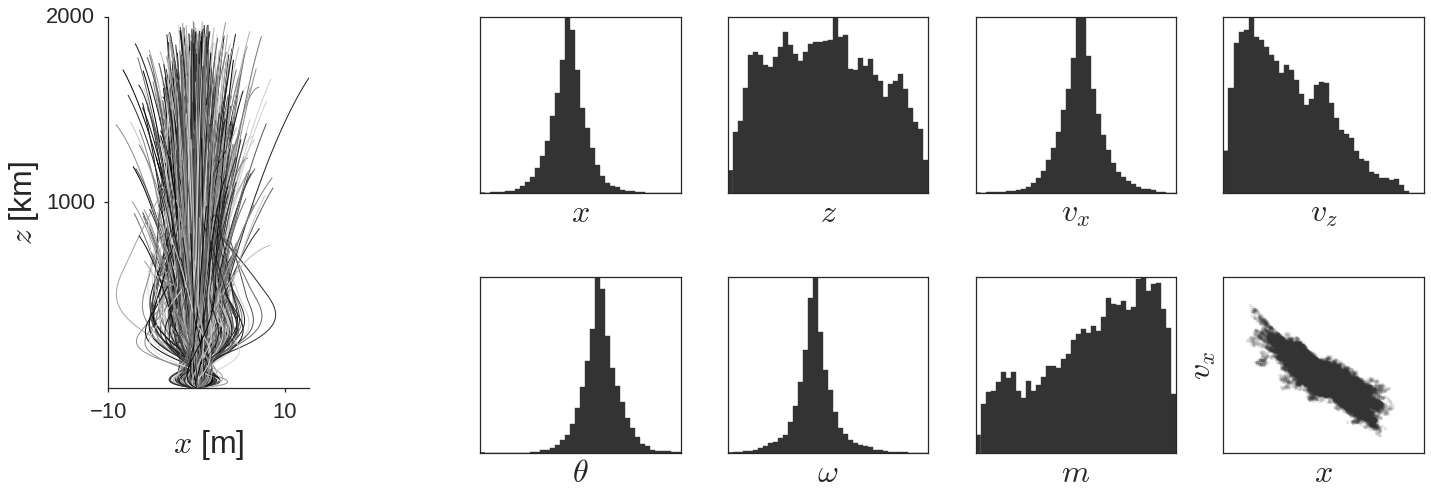} 
      \caption{Random trajectories, distribution of the initial states and joint distribution of $x$ and $v_x$ (TVR).}
      \label{fig:tv_data_generated} 
\end{figure}

Finding the optimal control for the Thrust Vectoring Rocket (TVR) from arbitrary initial conditions in $\mathcal A$ (see Table \ref{table:bounds}) revealed, instead, to be too challenging for the optimal control solver used here, thus the random walk is initialized around a perfect vertical landing scenario $x=0$ $v_x=0$ $\theta = 0$ $\omega = 0$ with the remaining variables being randomly initialized in $\mathcal A$. The random walk will then take care of continuing this trivial case into diverse initial conditions, but will not be able to fill $\mathcal A$ uniformly. Example of the optimal trajectories thus computed and the histograms of their initial states are shown in Figure \ref{fig:tv_data_generated}. We can see how, due to the repetition of the initial state, the distribution of some variables, particularly $x$, $v_x$, $\theta$ and $\omega$, approximates a Gaussian distribution around the nominal descent. In the same figure the joint distribution of $x$ and $v_x$ is included, showing an inverse relation between these variables that corresponds to trajectories roughly pointing to the landing position in the horizontal axis. Trajectories with a high initial $v_x$ pointing away from $x$ are thus not in the training dataset, which is not an issue since those cases are not expected in real landing scenarios.

\section{Learning the Optimal Control}
\label{sec:nnarchitecture}

Deep neural networks with a simple feed-forward architecture are trained to learn the optimal state action relation from the generated data. Eq.(\ref{eq:hjb2}) shows, mathematically, how the quantity to be learned is a function of the state alone (not his history), suggesting a feed-forward deep architecture is indeed an appropriate choice. As each model has several control variables, we train separate networks for each of them. The following section describes in more details the feed forward architectures considered and the training procedure.

\subsection{DNN architecture}

Architectures with different numbers of layers and units (neurons) per layer are considered. For comparison purposes, both shallow networks (one hidden and one output layer) and deep networks (several hidden layers) will be studied. The output $o_{ij}$ of the unit $i$ of layer $j$ can be expressed as:
\begin{equation}
o_{ij} = g(\mathbf{w_{ij}} \mathbf{o_{i-1}} + b_{ij}),
\end{equation}
where $\mathbf{w_{ij}}$ is the vector of weights and $b_{ij}$ is the bias corresponding to that unit, $ \mathbf{o_{i-1}}$ is the full output of the previous layer and $g$ is a non-linear function. 

The selection of the non-linearities $g$ (neuron types) has been identified as one of the most important factors of DNN architectures \cite{jarrett2009best} and thus different functions are considered for the hidden and output layers. For the hidden layers classical sigmoid units are compared to rectified linear units (ReLUs), which corresponds to the activation function $max(0,x)$. It has been pointed out that ReLu units have two main benefits when compared to sigmoid functions: they do not saturate, which avoids the units to stop learning in deep networks (the vanishing gradient problem), and the output of the units is frequently zero, which forces a sparse representation that is often addressed as a way of regularization that improves the generalization capabilities of the model \cite{glorot2011deep}. The sigmoid function used for the comparison is the hyperbolic tangent, selected based on their better convergence compared to the more common logistic function \cite{LeCun2012tanh}. 

\begin{figure}[t]
\centering
      \includegraphics[width=0.7\textwidth]{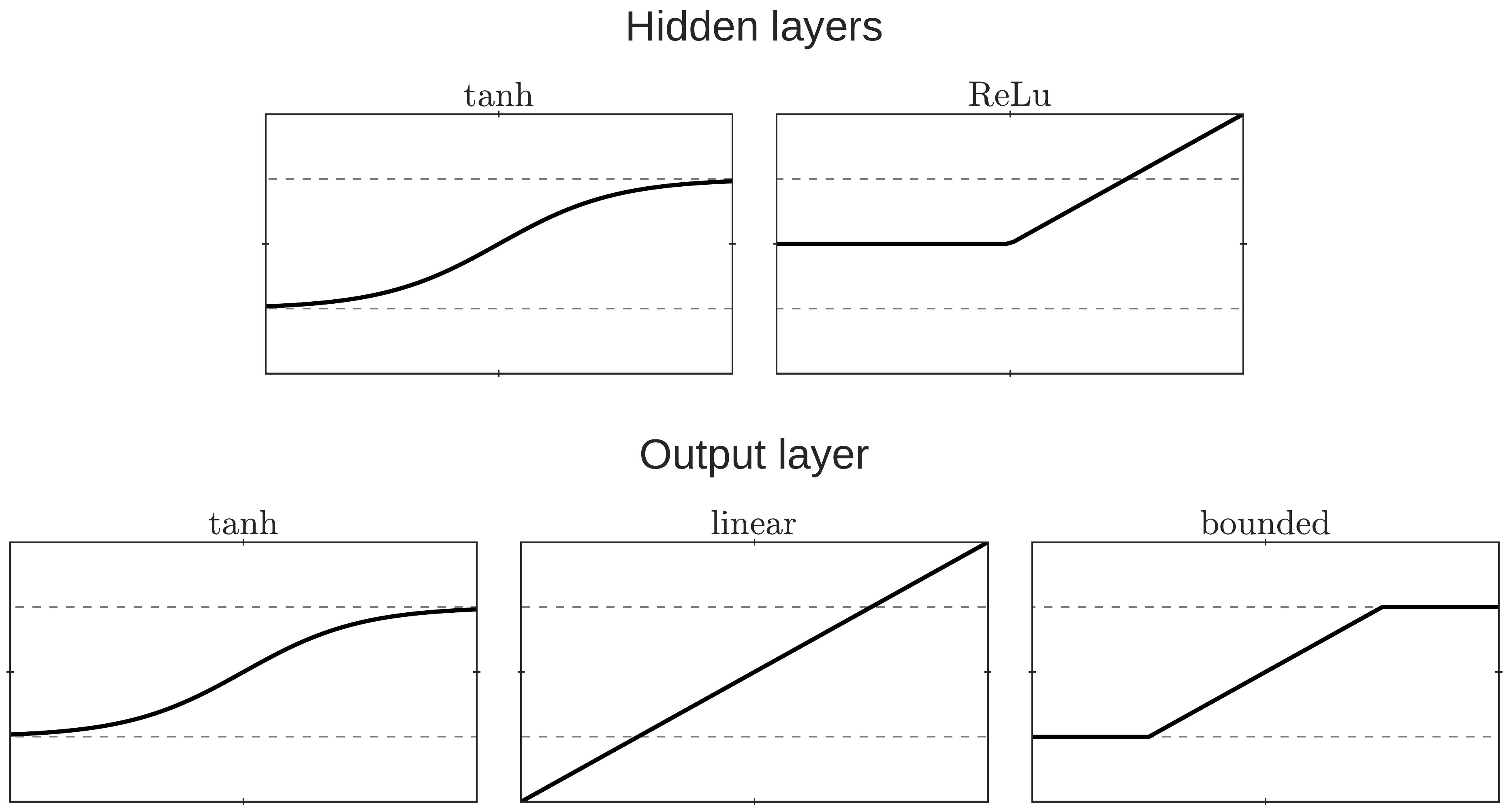} 
      \caption{Functions $g$ considered for the hidden and output layers.}
      \label{fig:nonlins} 
\end{figure}

For the output layer, the following functions $g$ are considered: the hyperbolic tangent function, the linear output $g(x) = x$ and a bounded linear output $g(x) = max(m,min(M,x))$, where $m, M$ are the bounds of the control variable. The bounded linear output function here propsed, unusual in the machine learning community, tries to leverage the fact that the optimal control is saturated in some of the problems (bang-bang structure). All the inputs and outputs of the network are normalized by subtracting the mean and dividing by the standard deviation (of the training data). Additionally, when using the hyperbolic tangent in the last layer, the normalized outputs are scaled to the range of the function ($[-1,1]$.). Figure~\ref{fig:nonlins} shows all the functions considered for the hidden and output layers.

\subsection{Training}

All networks are trained until convergence with stochastic gradient descent (SGD) and a batch size of $b=8$. At each iteration, the training process seeks to minimize the squared loss function $C=\sum_{i=0}^b \frac{1}{b} ( \mathcal N(\mathbf {x_i}) - y(\mathbf {x_i}))^2$ for the neural network output $\mathcal N(\mathbf{x_i})$ and the optimal action $y(\mathbf {x_i})$. Each weight $w$ is then updated with a learning rate $\eta=0.001$ and momentum with $\mu=0.9$ \cite{sutskever2013importance}: 

$$v_i \rightarrow v_i' = \mu v_i - \eta\frac{\partial C}{\partial w_i}$$
$$w_i \rightarrow w_i'= w_i + v_i'$$

After every epoch, the loss error is computed for a small portion of training data ($5\%$) that is not used during training, and an early stopping criteria based on the idea of \textit{patience} \cite{bengio2012practical} is used to determine when to stop. In essence, all networks are trained for $p$ more epochs after the last epoch that showed an improvement. The \textit{patience increment} $p$ is here set to 5.

Xavier's initialization method \cite{glorot2010understanding} is used to randomly set the initial weights. Although it was designed to improve the learning process for logistic units, it has been shown that this idea can also be beneficial for networks with ReLu units \cite{he2015delving}. Each weight $w_i$ is drawn from a uniform distribution $U[-a,a]$, with $a=\sqrt{\frac{\beta}{\text{fan}_\text{in}+\text{fan}_\text{out}}}$, where $\beta=12$ for the ReLu units and $\beta=6$ for the linear and $tanh$ units, and $\text{fan}_\text{in}$, $\text{fan}_\text{out}$ are the number of units of the previous and following layer.

\section{Evaluation}
\label{sec:evaluation}

Given any state of an optimal landing trajectory, the mean absolute error (MAE) is used to evaluate the difference between the optimal actions and the network predictions. This measure allows the comparison of the performance of different DNNs but does not provide an accurate measurement of how well the landing task is accomplished. Small errors could be propagated through the trajectory resulting in sub-optimal or failed landings. Errors could even potentially be corrected without impacting on the landing success. We thus need an added evaluation scheme to be able to judge how well a DNN has accomplished to learn the optimal control of the given landing scenario. For this purpose we introduce the DNN-driven trajectories: simulations of the landing dynamics as controlled by the DNN.

The DNN-driven trajectories are computed by numerical integration of the system dynamics $ \mathbf {\dot x} = \mathbf f(\mathbf x, \mathbf u)= \mathbf  f(\mathbf x,\mathcal N(\mathbf x))$:

$$\mathbf x(t) = \int_{0}^{t}\mathbf  f(\mathbf x(\tau),\mathcal N(\mathbf x(\tau)))d\tau$$

The DNN-driven trajectory will reach the target point $\mathbf x_t$ with some error, thus a tolerance region is defined around $x_t$ and a trajectory is considered successful when it reaches this region. The tolerance $\tau$ for each problem is defined for the final value of each state variable (position $\tau_r$, velocity $\tau_v$, angle $\tau_{\theta}$ and angular velocity $\tau_{v\theta}$) and can be found in table \ref{table:tolerances}. These tolerance values are between 0.5\% and 1\% of the range of each variable in the training set. 

The final state $\mathbf x_f$ of a DNN-driven trajectory is defined as the closest state to the target: $\mathbf x_f  = \argmin_{x_i} |\mathbf x_t - \mathbf x_i|$. Given that the state includes variables with heterogeneous units, when computing  $|\mathbf x_t - \mathbf x_i|$ we use $\tau_i$ as units (ie. a final distance of $\tau_r$ m is deemed as equivalent to a final velocity of $\tau_v$ m/s).

\begin{table}[]
\centering
\caption{Range of the variables and value of the crash tolerance $\tau$ (roughly $0.5-1$\% of the range).  \label{table:tolerances}}

\begin{ruledtabular}
\begin{tabular}{rcccccccc}
               & \multicolumn{3}{c}{\textbf{Quadcopter}} && \multicolumn{4}{c}{\textbf{Simple Spacecraft}}   \\                                                      
\textbf{}      & r                            & v                             & $\theta$   &&& r                           & v                             &                             \\ \cline{2 - 4} \cline{7- 8} 
range:         & 20.62 m & 17.34 m/s & 100.84$^{\circ}$ &&& 2018.0 & 135.4 & \\
$\mathbf \tau:$ & 0.1 m                         & 0.1 m/s                         & 1$^{\circ}$ &&& 10 m                         & 0.7 m/s                         &         \\
\\
               & \multicolumn{3}{c}{\textbf{Reaction Wheel Spacecraft}} && \multicolumn{4}{c}{\textbf{Thrust Vectoring Rocket}}   \\                                                      
\textbf{}      & r                            & v                             & $\theta$   && r                            & v                             & $\theta$   & $v_ \theta$                         \\ \cline{2 - 4} \cline{6- 9} 
range:         & 2022.3 m & 141.6 m/s & 80.2 $^{\circ}$ && 1991.4 m & 73.88 m/s & 2.23 $^{\circ}$ & 0.40 $^{\circ}$/s\\
$\mathbf \tau:$ & 10 m                         & 0.7 m/s                         & 1$^{\circ}$ && 10 m                         & 0.7 m/s                         & 0.02 $^{\circ}$   &  0.004 $^{\circ}/s$  \\

\end{tabular}

\end{ruledtabular}

\end{table}
To evaluate the performance of the DNNs in terms of optimality, it would seem obvious to compare the cost function $J(\mathbf x_0)$ evaluated at $\mathbf x_f$ along a DNN-driven trajectory, to the optimal cost $J^*(\mathbf x_0)$. However, due to the introduced tolerances, said cost function can result to be slightly better than the optimal cost. To get a fairer comparison, the optimal cost is also computed stopping the optimal trajectory at the same distance to the target as $\mathbf x_f$.

The final evaluation of a DNN-driven trajectory is thus fully described by several quantities: the success rate SR (i.e. the likelihood to actually get to the target point within the set tolerances), the distance of $\mathbf x_f$ to the target value in terms of $\mathbf r$, $\mathbf v$ and, according to the model, $\theta$ and $\omega$, and, in case the DNN-trajectory is deemed as successful, the optimality defined as the relative error of the cost function $J$ with respect to the value of optimal control solution. 

\begin{figure}[t]
\centering
      \includegraphics[width=1\textwidth]{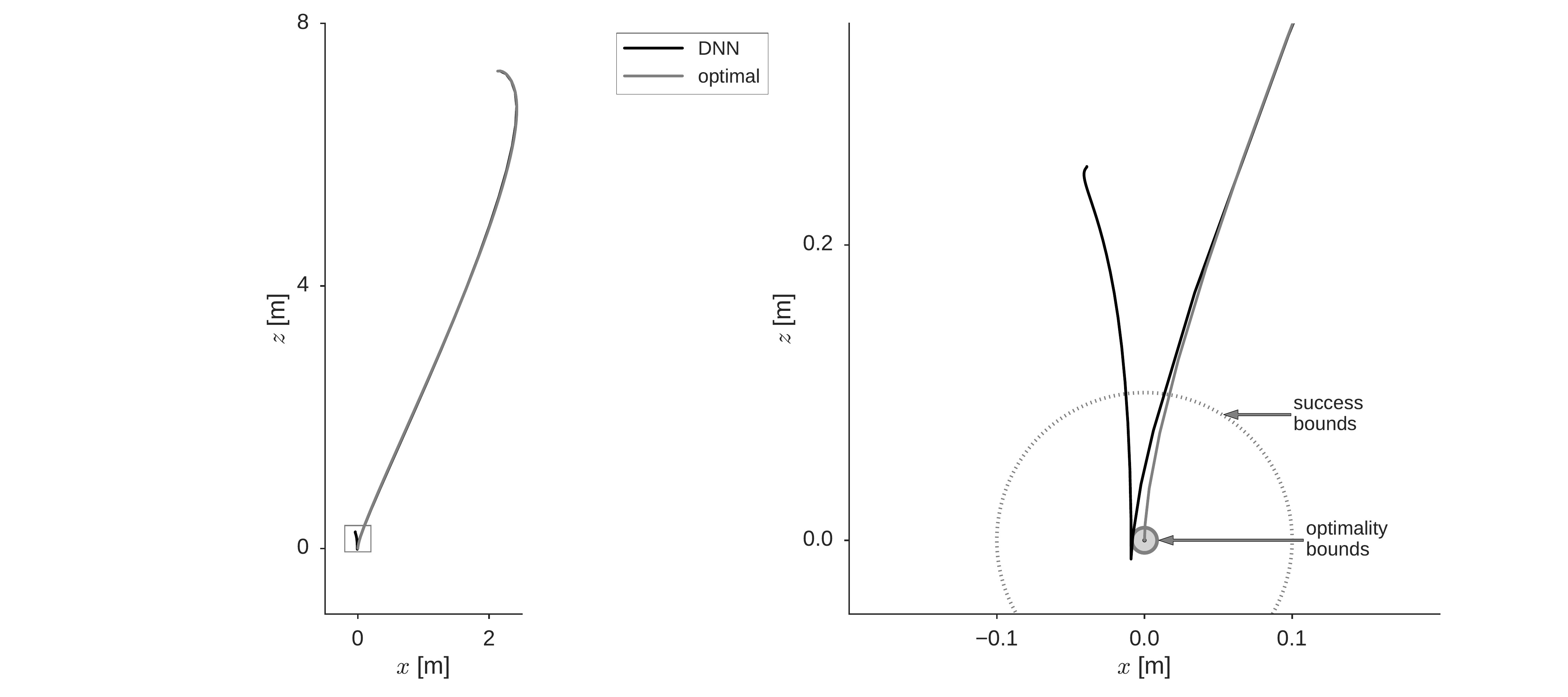} 
      \caption{Last part (right) of a QUAD-QC landing trajectory (left) depicting the \textit{success} and \textit{optimality bounds}.      \label{fig:target} }
\end{figure}

Figure \ref{fig:target} shows an example landing trajectory where the \textit{success bounds} indicate the tolerance around $x_t$ and the \textit{optimality bounds} show the points used to compare the DNN-driven trajectory to the optimal solution. The optimality of the trajectory is evaluated up to the points where the optimal and NN-Driven trajectories intersect with the \textit{optimality bounds}. In the figure, only the distance to the goal has been considered (not the velocity or the angle).

\section{Results}
\label{sec:results}

The Mean Absolute Error computed for the states of the optimal trajectories in the validation set is used in order to compare the architectures of various artificial neural networks with different number of layers and different non-linearities. To provide an analysis of the different architectures, four control variables representing the different control profiles (as illustrated in Figure \ref{fig:control_profiles}) are selected:

\begin{itemize}
    \item S-MOC ($u_2$): smooth and continuous
    \item Q-QC ($u_2$): smooth and continuous with saturated regions 
    \item RW-MOC ($u_1$): bang-bang (always saturated)
    \item RW-MOC ($u_2$): continuous with abundant plateaus and sharp transitions
\end{itemize}

Then, in order to study the performance of the trajectories produced by the DNNs for each problem, trajectories from 1,000 different initial states are simulated and evaluated.

\subsection{Neuron type: non-linearity selection}

Table~\ref{table:nonlinearities} shows the results of the evaluation for the four selected control variables. The Mean Absolute Error (MAE) is computed for deep (5 layers) networks with 32 units per layer and different neuron types for the hidden and output layers on the test set.

\begin{table}[]
\centering
\caption{Effect of the neuron type: MAE of DNNs with 3 layers and 32 units/layer}
\label{table:nonlinearities}
\begin{ruledtabular}

\begin{tabular}{ccccccc}

&  \textbf{t-t} & \textbf{t-l}  & \textbf{t-b} & \textbf{R-t} & \textbf{R-l}  & \textbf{R-b} \\     \cline{2-7}

\textbf{SSC-MOC ($u_2$)}      & 0,0257               & 0,0339               & 0,0382               & \textbf{0,0227}      & 0,0260               & 0,0259               \\

\textbf{QUAD-QC ($u_2$)}  & 0,0697               & 0,0537               & 0,0668               & 0,0345               & 0,0371               & \textbf{0,0321}      \\

\textbf{RW-MOC ($u_1$)}                   & 651,5             & 887,9             & 646,6             & 458,3            & 671,4             & \textbf{304,8}    \\

\textbf{RW-MOC ($u_2$)}  & \textbf{0,000934}    & 0,000981            & 0,000952            & 0,00114            &  0,00105              & 0,00108             \\
\end{tabular}
\end{ruledtabular}
\raggedright{\textit{hidden-output neuron type: t ($\tanh$), R ($\textit{ReLu}$), l ($\textit{lin}$), b ($\textit{bounded}$)}}
\end{table}

In the case of the two control variables with saturated regions ($u_2$  in QUAD-QC and $u_1$ in RW-MOC) the best results are provided by networks with a bounded linear output, closely followed by those obtained by networks with a $\tanh$ non-linearity on the output layer. These two networks can easily reproduce the saturated regions: the bounded output by producing values higher or lower than the saturation level and the $\tanh$ by producing values as high or low as possible. Networks with linear outputs, however, correspond to the lowest performance on these cases, as expected given that these network need to output the exact saturation value.

Regarding the neurons in the hidden layers, the DNNs with $\textit{ReLu}$ units outperform those with $\tanh$ in most cases. Better results are consistently obtained for $\textit{ReLu}$ units in the first three models ($u_2$ in SSC-MOC, QUAD-QC and $u_1$ in RW-MOC) when compared to networks with the same output but $tanh$ hidden units. This difference is particularly large in the two models with bounded profiles, where the performance of the ReLu units is up two two times better. An exception to this is the variable $u_2$ of RW-MOC, where networks with $\tanh$ units consistently perform better, although the difference, in this case, is small. Modelling the abundant plateaus present in this profile could be challenging for networks with ReLu units, while the flat areas of the $tanh$ function could be an advantage for this problem. In any case, it is not possible to conclude that a nonlinearity is better across all domains and the difference between them should be addressed for new models. 
The rest of the evaluation in this section is be done using ReLu units for the hidden layers and $tanh$ and bounded linear functions for the output layers depending on the type of profile, being the later used for profiles including saturation.

\subsection{Depth of the network}

The mean absolute error (MAE) is computed for models ranging from 2 to 5 hidden layers and different numbers of units per layer. The results are included in Table \ref{table:depth}, note that networks with more layers always outperform shallower networks with a similar number of parameters. DNNs with 5 layers and just 16 units per layer are consistently better than shallow networks with 515 units in the hidden layer, even when the latter has almost 4 times more parameters. DNNs with 5 layers and 32 units will be used in the following sections.

\begin{table}[tb]
\centering
\caption{Mean Absolute Error (MAE) of networks with different numbers of layers and units)}
\label{table:depth}
\begin{ruledtabular}

\begin{tabular}{cccccc}

\multirow{2}{*}{\textbf{layers-units}}
& \multirow{2}{*}{\textbf{\#weights${}^{*}$}}
&  \textbf{QUAD-QC}
&  \textbf{SSC-MOC}
&  \textbf{RW-MOC}
&  \textbf{RW-MOC} \\

&
&\textbf{($u_2$)}
&\textbf{($u_2$)}
&\textbf{($u_1$)}
&\textbf{($u_2$)} \\

2 - 256 &  1,793 & 0,0580 & 0,0357 &   752,4  & 0,00188 \\
2 - 512 &  3,585 & 0,0577 & 0,0297   & 618,1 & 0,00150  \\
\\
3 - 16 &  385 &  0,0625 & 0,0341  & 677,4 & 0,00201 \\ 
3 - 32 &  1,281 & 0,0524 & 0,0330   & 551,1 & 0,00138  \\
3 - 64 &  4,609 & 0,0436    &  0,0257 &  497,2 & 0,00123  \\
\\
4 - 16 &  657 &    0,0503   &      0,0271 & 568,8  & 0,00161  \\
4 - 32 &  2,337 &   0,0480  &  0,0250 & 592,0 & 0,00121 \\
\\
5 - 16 &  929 &     0,0475  &   0,0208  & 474,8 & 0,00148 \\
5 - 32 &  3,393 &     0,0321  &   0,0227  & 304,8 & 0,00114 \\

\end{tabular}
\end{ruledtabular}
\raggedright{\textit{${}^{*}$ Number of weights for a network with 5 inputs (QUAD and SSC), for RWSC (6 inputs), the number of weights is increased by the number of units per layer.}}
\end{table}

\subsection{DNN-driven trajectories}

The DNN-driven trajectories are evaluated as described in section \ref{sec:evaluation}. A summary of the results is included in table \ref{table:trajectory_evaluation}. High success rates are achieved across all domains while obtaining a low relative error with respect to the value of the optimal trajectories.

\begin{table}[]
\centering
\caption{Performance of the DNN-driven trajectories. \label{table:trajectory_evaluation}}

\begin{ruledtabular}
\begin{tabular}{lcccccc}
&\textbf{Success rate }      &  \multicolumn{4}{c}{\textbf{Distance to target}} &  \textbf{Optimality}             \\  \cline{3-6}
&& $r$ [m] & $v$ [m/s] & $\theta$ [deg] & $\omega$  [deg/s] \\

\textbf{QUAD-QC}  & 100.0\% & 0.014& 0.027& 0.36 & - & 1.82\% \\ 
\textbf{QUAD-TOC} & 100.0\% & 0.016& 0.028& 0.48 & - & 1.12\% \\ 
\\
\textbf{SSC-QC}  & 100.0\% & 0.40 & 0.052 & -& -&  0.24\% \\ 
\textbf{SSC-MOC}  & 100.0\% & 2.47 & 0.12 & -& -&  0.45\% \\ 
\\
\textbf{RWSC-QC}  & 100.0\% & 0.29 & 0.044 & 0.20 & -& 0.40\% \\ 
\textbf{RWSC-MOC}  & 98.3\% & 2.90 & 0.073   & 0.31 & - &  0.72\% \\ 
\\
\textbf{TVR-QC}  & 99.0\% & 1.10 & 0.066 & 0.06 & 0.0075&  0.38\% \\ 
\textbf{TVR-MOC}  & 95.0\% & 1.95 & 0.094   & 0.012 &  0.0054 & 0.33\% \\ 
\end{tabular}
\end{ruledtabular}
\end{table}

The quadcopter model and the simple spacecraft achieve a $100\%$ success rate for the two objective functions in each case. In all cases the distance to the target $D(\mathbf x_f)$ is way below the success bounds showing that the state $\mathbf x_f$ reached by the network is close to the target state $\mathbf x_t$. The relative error of $J$, lower than $2\%$ for the quadrotor problems and lower than $0.5\%$ for the simple model, indicates that the profile followed by the network accurately represents the optimal control.

\begin{figure}[tb]
\centering
      \includegraphics[width=0.8\textwidth]{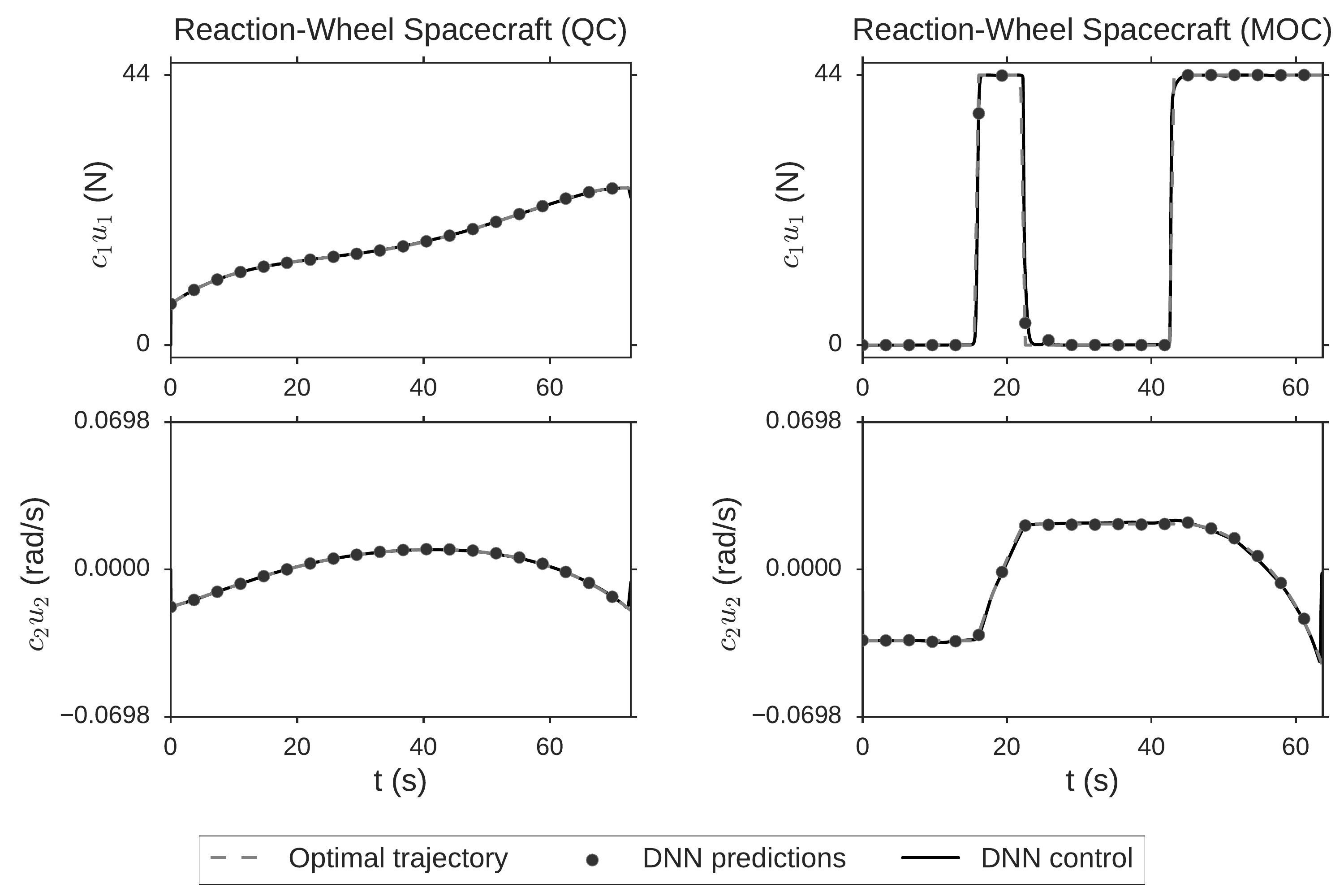}
      \caption{DNN predictions during the optimal trajectory and during a DNN-driven trajectory (RWSC problems)}
      \label{fig:rw_controls} 
\end{figure}

Similar results are obtained for the case of the reaction-wheel spacecraft model, with $100\%$ and $98.3\%$ success rates for quadratic and time optimal control. A low distance to the target is obtained and the relative error of $J$ is below $1\%$ for both objective functions. Figure~\ref{fig:rw_controls} shows an example of the DNN predictions and the trajectory driven by the DNN. It is interesting to note that the predictions are accurate even for the case of $u_2$ in the RWSC-MOC where the numerous plateaus were expected to be difficult to approximate with the neural networks. Although this figure is only included as an example of the DNN predictions and DNN-driven trajectories, similar results are obtained for the other problems here considered.

\begin{figure}[t]
\centering
      \includegraphics[width=1\textwidth]{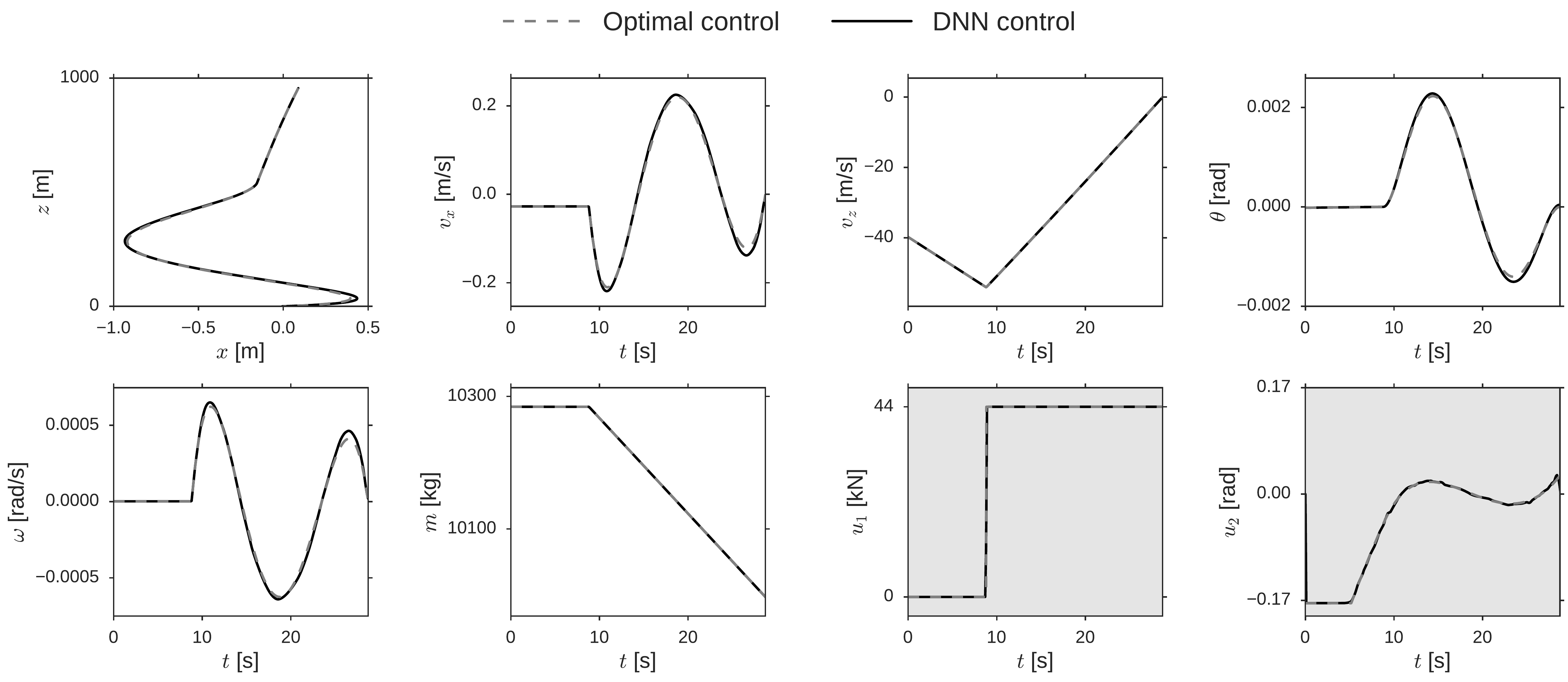}
      \caption{State and control variables during a DNN-driven landing  (TVR-MOC problem).}
      \label{fig:tv_moc_all_vars} 
\end{figure}

Slightly lower success rates are achieved for the thrust vectoring (TVR) models ($99.0\%$ and $95.0\%$ for the QC and MOC objectives), but the DNN is still able to reproduce the controls as illustrated by Figure~\ref{fig:tv_moc_all_vars}, where all the state and control variables of a TV-MOC landing are included.
Similarly to the previous cases, the relative error of $J$ is below $0.5\%$ for both objective functions.

\clearpage

\subsection{Behaviour after reaching $x_f$}
\label{sec:after_target}

The ideal behaviour of a DNN-driven trajectory after it reaches the final point $x_f$ would be to start a hovering phase at the exact target position $x_t$. However, in all the considered cases, the structure of the optimal control makes it impossible to learn such a behaviour for the DNNs. This is clear, for example, in the cases of time and mass optimal control, where the target position is always reached with either maximum or minimum thrust, but a thrust exactly equal to $mg$ is required for hovering at that position, a value that will never be present in the data set and is thus difficult to learn. Similarly, in the other models, the target state $x_t$ is reached with different control values $u$ that do not necessarily correspond to the value required for hovering and thus making it impossible for the network to learn how to reach and hover the exact $x_t$ position. Bearing in mind the impossibility for a DNN to learn how to hover at $x_t$, we analyse the behaviour of a DNN-driven trajectory after it reaches its final target point $x_f$. 

\begin{figure}[t]
\centering
      \includegraphics[width=1\textwidth]{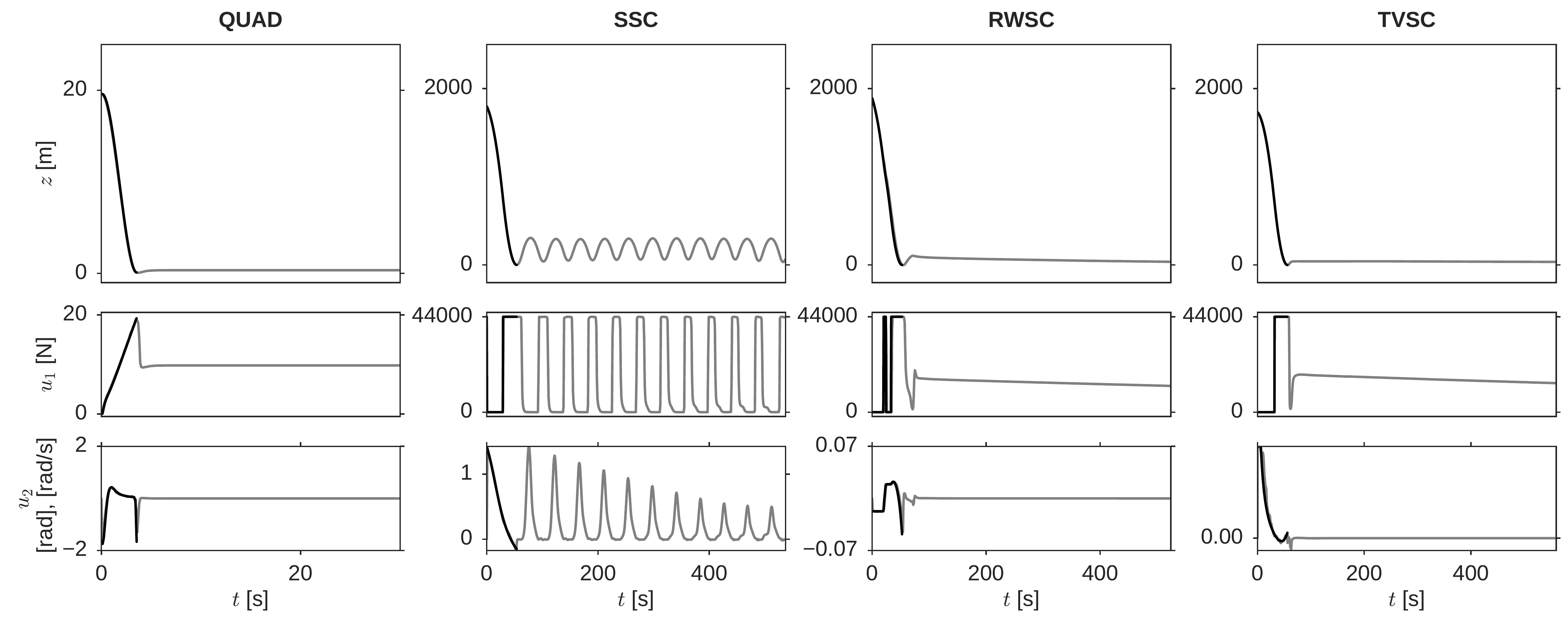}
      \caption{ Altitude and control variables of four models until $x_f$ is reached (black) and afterwards (gray). }
      \label{fig:after_target} 
\end{figure}

Figure \ref{fig:after_target} shows DNN-driven trajectory  behaviour after reaching $x_f$ for the quadcopter (quadratic control) and the other models (mass optimal control). Remarkably, in the case of the QUAD, RWSC and TVR dynamics, a hovering behaviour is observed at a position close to $x_t$. For the QUAD case the numerical value of the thrust $c_1u_1$ at this point acquires an approximate value of $9.81$ [m/s$^2$] even for the case of time optimal control when only saturated values are present in the training data. For the TVR and RWSC case the numerical value of $c_1u_1$ is, instead constantly decreasing in time as the mass of the rocket is also decreasing. 

The behaviour is different for the model where the pitch angle is directly controlled (SSC). In this case there is no hovering phase, but the spacecraft continuously oscillates above the target position. This is not unexpected, as the optimal value of $u_2 = \theta$ at the final target point is not unique and the DNN can only learn its average value which will likely be different from zero. As a consequence, the spacecraft thrust is not vertical around the target position, excluding the possibility to hover.

\subsection{Generalization}

It is of interest to study the behaviour of DNN-driven trajectories when the initial conditions are outside of the initialization area $\mathcal A$. If the networks have learned an approximation to the solution of the HJB equations, we would expect them to still be able to represent the action for states that are not included in the training data. The results for 1000 DNN-driven trajectories starting from initial conditions drawn from two different extensions of $\mathcal A$ are shown, each one excluding the previous area. The quadcopter (QC), the simple and reaction-wheel spacecrafts (MOC) are considered. A summary of the results is included in Table~\ref{table:generalization_evaluation}.

\begin{table}[]
\centering
\caption{ Performance of the DNN-driven trajectories outside of $\mathcal A$       \label{table:generalization_evaluation}}

\begin{ruledtabular}

\begin{tabular}{ccccccc}
  &  \textbf{Success Rate}  &  \multicolumn{3}{c}{\textbf{Distance to target}} & &\textbf{Optimality error}         \\  \cline{4-6}
&&& $r$ [m] & $v$  [m/s] & $\theta$ [deg] & \\

& \textbf{ $\mathcal A_{~}$} & 100\%  & 0.014& 0.027& 0.36 & 1.82\% \\ 
\textbf{QUAD-QC} &\textbf{$\mathcal A_1$}& 84.4\% & 0.036& 0.077& 0.34 & 3.53\% \\ 
& \textbf{$\mathcal A_2$}  & 75.0\% &  0.29& 0.21& 1.63 & 5.59\% \\ 
\\
&\textbf{$\mathcal A_{~}$}   & 100\%  & 0.40 & 0.052 &  - & 0.24\% \\ 
\textbf{SSC-MOC} &\textbf{$\mathcal A_1$} & 88.8 \% &  27.71 & 0.29 &  - &  0.64\% \\ 
&\textbf{$\mathcal A_2$}  & 57.8 \% & 521.44 & 1.741 &  - &  1.31\% \\ 
\\
&\textbf{$\mathcal A_{~}$}  & 98.3\% & 2.90 & 0.073   & 0.31 &  0.72\% \\ 
\textbf{RWSC-MOC}&\textbf{$\mathcal A_1$} & 57.9\% &  9.34 & 0.28   & 0.35 & 1.13\% \\ 
&\textbf{$\mathcal A_2$} & 34.9\% & 9.48& 1.72 & 0.58 & 0.86\% \\ 
\end{tabular}

\end{ruledtabular}
\end{table}

\begin{figure}[t]
\centering
      \includegraphics[width=1\textwidth]{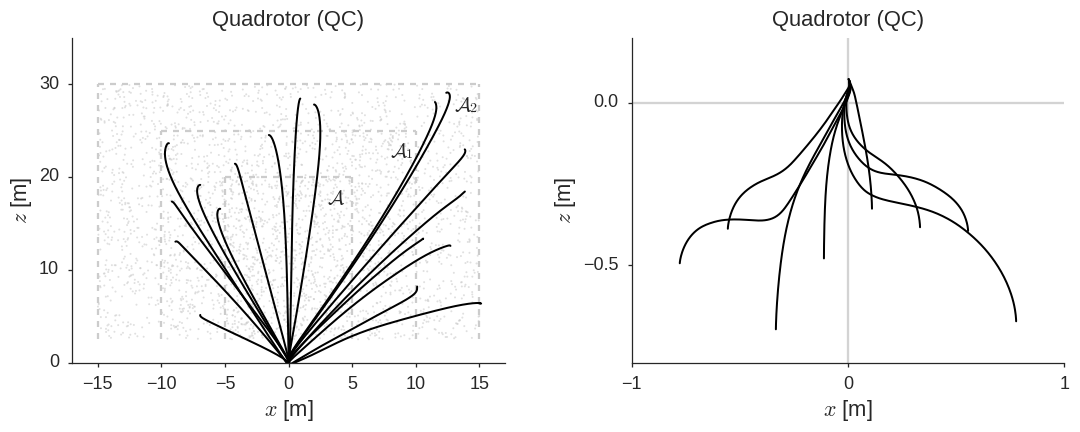}
      \caption{Generalization of the DNN (QUAD-QC) to extensions of $\mathcal{A}$ and initial states with $z=0$.}
      \label{fig:generalization_quad_qc} 
\end{figure}

In the case of the quadrotor, the trajectories are selected from an extension $\mathcal A_1$ of $5$ [m] both in $x$ and $z$ and an extension $A_2$ of $10$ [m]. Success rates of $84.4\%$ and $75.0\%$ are obtained for these extensions, although the average distance to the target is still close to 0, being of $0.29$ [m], $0.21$ [m/s] and $1.63$ [$^\circ$] for the furthest extension $A_2$. The optimality error in the extensions is $3.53$ and $5.59\%$. Figure~\ref{fig:generalization_quad_qc} shows some examples of these trajectories. Remarkably, the ability of the network to achieve the final target extends also for initial conditions lower than the landing position as illustrated in the same figure, which requires thrusting to move upwards, a condition not encountered during training. 

\begin{figure}[t]
\centering
      \includegraphics[width=1\textwidth]{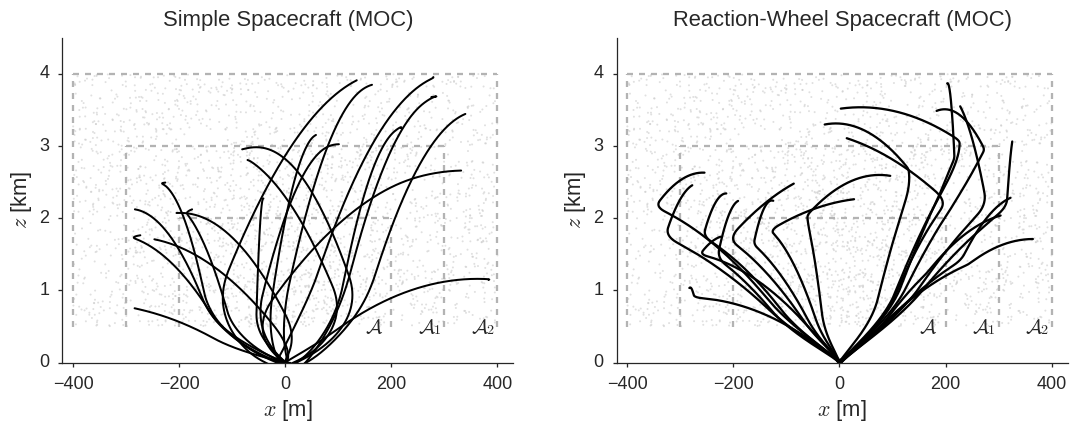}
      \caption{Generalization of the DNN (SSC-MOC and RWSC-MOC) to extensions of $\mathcal{A}$}
      \label{fig:generalization_spacecrafts} 
\end{figure}

For the spacecraft models, the $x, y$ coordinates of the possible initial conditions are extended by $100$, $1000$
[m] for ($\mathcal A_1$) and  $200, 2000$ [m] for $\mathcal A_2$. 
The simple spacecraft achieves success rates of $88.8\%$ and $57.8\%$, but very high distances to the target are obtained, as some trajectories result in catastrophic trajectories ending up far from $x_t$. The trajectories obtained for the reaction wheel model, although achieving lower success rates, have a lower average distance to the target, as points close to the target are always reached. Figure~\ref{fig:generalization_spacecrafts} shows some examples of trajectories for these models.

\begin{figure}[t]
\centering
      \includegraphics[width=1\textwidth]{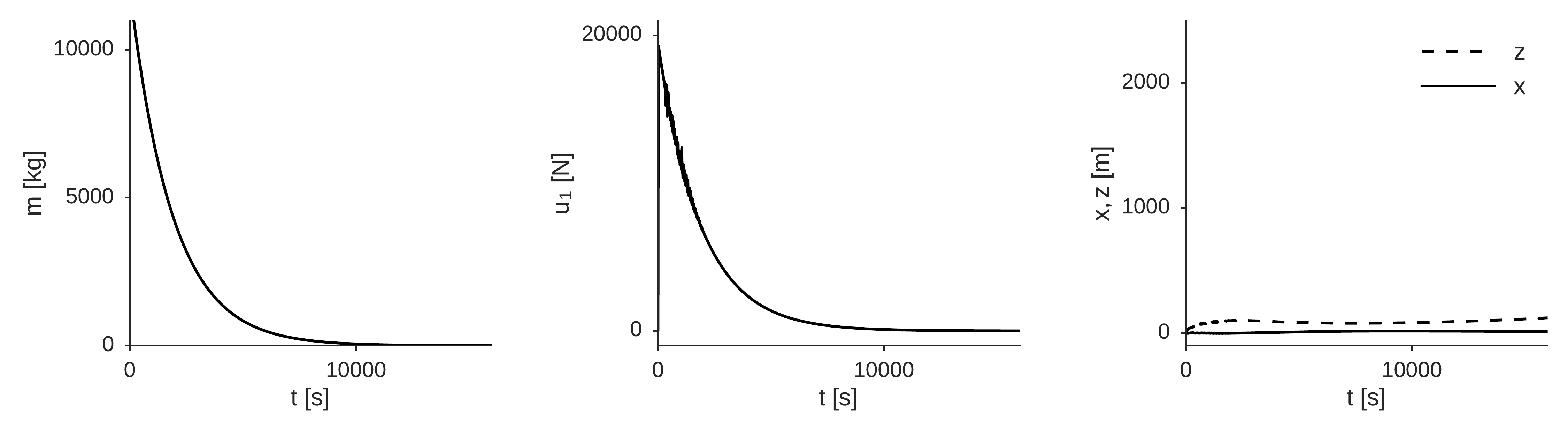} 
      \caption{RW-MOC landing with the long after the final state $t_f$ has been reached.}
      \label{fig:mass_after} 
\end{figure}

A further, intriguing, property of the DNN-driven trajectories is revealed observing the RWSC and TVR models behaviours for a long time after they reach the target point. As previously noted a hovering behaviour is observed and it persists in time long after the acquisition of the target state. The reduction of the mass due to the propellant loss is compensated by a commanded reduction of the thrust ($c_1 u_1$) as illustrated in Figure~\ref{fig:mass_after}. The spacecraft hovers close to the target state long after reaching $x_f$, even when the mass of the spacecraft is reduced to a fraction of the values found on the training data. It is thus clear how the network has learned in some way the problem dynamics and exploits it to maintain the spacecraft close to the target position.

\section{Conclusion}

In this paper it has been shown how deep neural networks (DNN) can be trained to learn the optimal
state-feedback in a number of continuous time, deterministic, non-linear systems of interest in the aerospace
domain. The trained networks are not limited to predict the optimal state-feedback from points within the subset
of the state space used during training, but are able to generalize to points well outside the training data,
suggesting that the solution to Hamilton-Jacobi-Bellman (HJB) equations is the underlying model being learned.
The depth of the networks has a great influence on the obtained results, being remarkable that shallow networks,
while trying to approximate the optimal state-feedback, are unable to learn its complex structure
satisfactorily. 
Our work opens to the possibility to design real-time optimal control architectures for planetary landing using
a DNN to drive directly the state-action selection. With this respect, the error introduced by the use of the
trained DNN not only does not have a significant impact on the final cost function achieved, but it is also safe
in terms of avoiding catastrophic failures for conditions that are far from nominal.

%



\section*{References}
\bibliographystyle{aiaa}

\bibliography{bib}
\end{document}